\documentclass[12pt]{article}
\usepackage{amssymb}
\usepackage{amsmath}
\usepackage{color}
\usepackage{graphicx}

\usepackage{hyperref}
\usepackage{bbm}
\usepackage{authblk}

\newcommand{\tr}{\mathrm{Tr}}

\title{Spinning Fluids: A Group Theoretical Approach}

\author[1]{Dario Capasso\footnote{dcapass00@ccny.cuny.edu}}
\author[2,3]{Debajyoti Sarkar\footnote{dsarkar@gc.cuny.edu}}
\affil[1]{Physics Department, City College of New York - CUNY, New York, NY 10031}
\affil[2]{Department of Physics and Astronomy, Lehman College of the CUNY, Bronx NY 10468, USA}
\affil[3]{Graduate School and University Center, City University of New York, New York 10016, USA}

\date{}

\begin{document}

\maketitle
%\hfill\raisebox{300pt}[0pt][0pt]{CCNY-HEP-12/10}
\begin{abstract}
The aim of this article is to introduce a Lagrangian formulation of relativistic non-abelian spinning fluids in group theory language. The corresponding Mathisson-Papapetrou equation for spinning fluids in terms of the reduction limit of de Sitter group has been proposed. The equation we find, correctly boils down to the one for non-spinning fluids. Two alternative approaches based on a group theoretical formulation of particles dynamics are also explored.
\end{abstract}

\tableofcontents

\section{Introduction}
\label{sec:intro}

In this article we discuss and generalize relativistic fluid dynamics from the alternate view point of group theory. This group theoretical approach was developed in \cite{Bistrovic:2002jx}. For a recent review see \cite{Jackiw:2004nm}. This technique relies on symmetry principles which makes it efficient to answer questions like anomalies etc. as recently discussed in \cite{Nair:2011mk}. In \cite{Nair:2011mk}, the technique used here to describe relativistic fluids was applied mainly to describe the Quark-Gluon Plasma (QGP), indeed QGP describes a fluid at strong coupling regime of QCD where the usual perturbative description of field theories fail; moreover now there is considerable evidence that QGP has several properties of a fluid. A different attempt is due to AdS/CFT correspondence which naturally describes strongly coupled field theories. 

The formalisms to describe particle motion in group theoretical terms was known from a long time (for some earlier references see \cite{Balachandran:1977ub}). The motivation for this formulation is not hard to understand either. To describe particles or fluids in a general spacetime, it's natural to look at the symmetries that (quantum) gravity offers. Though the natural selection is the Poincar\'e group (section~\ref{sec:P_Spinning_fluids}), the incorporation of cosmological constant formally calls for the de-Sitter group in 4 dimensions. The formalism in one word gives a way to couple the particle/fluid with these gauge groups. In this formalism it also becomes easier to introduce fluid vorticity.

The plan of the paper is as follows. For the rest of this introduction we discuss some models to describe relativistic abelian spinless fluids in the group theoretical language as was briefly addressed in \cite{Nair:2011mk}.  In section~[\ref{sec:dS_Spinning_fluids}] we describe the main model for describing relativistic spinning fluids based on a reduction of the de Sitter group. This is the main result of our paper. Later we present some alternate ways to formulate the problem and discuss each of their advantages and disadvantages. In section~\ref{sec:dS_torsion} we use a similar procedure using dS group, but in the context of torsion. Although torsion couples naturally with the spin density it is a property of spacetime that has not been observed; nevertheless this case should not be ignored. The model based on the Poincar\'e  group described in section~\ref{sec:P_Spinning_fluids} shows drawbacks over the others which come in the form of extra constraints that are not present in the other models. The fact that we can have several models is the generality of this approach. 
%We use the formalism in situations interesting for astrophysical observations. For example in section~\ref{section:FRW} we describe the fluids in an expanding FRW spacetime. On the other hand section~\ref{sec:Kerr} is devoted to the study of dust in the spacetime of a Kerr black hole. In this context we discuss some prospective experimental signatures and finally in appendix \ref{ap:Keplerian} we describe their relation with Keplerian orbits. Appendix \ref{appendixC} collects some well known properties of dS group generators according to our convention. %Finally we discuss some well known group properties in the appendix \ref{appendixC} that we used in our calculation. \textcolor{red}{Also in appendix \ref{ap:Keplerian} we plot the galaxy rotation curve which we get from our calculation of the Kerr black hole case. We end with some discussions of experimental signatures and outlooks.}

%%%%%%%%%%%%%%%%%%%%%%%%%%%%%%%%%%%
\subsection{Ordinary Abelian Fluid}\label{sec:ordinary_abelian_fluids}
%%%%%%%%%%%%%%%%%%%%%%%%%%%%%%%%%%%

%We start with the classical well-known hydrodynamical equations and provide an action. We recall how can they be written in terms of abelian group theory language. We generalize this group theoretic action for relativistic fluids and finally to fluids with non-abelian gauge groups. 

In 3+1 dimensions (with a mostly minus metric signature), an ordinary abelian fluid is described (in terms of collective variables and not as many particle motion) by the \emph{Continuity Equation}
\begin{eqnarray*}\label{conteq}
\partial_t\rho(t,\mathbf{r})+\nabla(\rho\mathbf{v}(t,\mathbf{r}))=0
\end{eqnarray*}
and \emph{Euler's eqauation}
\begin{eqnarray*}
\partial_t\mathbf{v}+(\mathbf{v}\cdot\nabla)\mathbf{v}=\mbox{\textbf{force}}
\end{eqnarray*}
where the \textbf{force} is the force acting on the unit volume of the fluid. $\rho$ and $\mathbf{v}$ are the fluid density and the velocity. These classical equations can also be found using a Lagrangian formulation with a Lagrangian \cite{Bistrovic:2002jx}
\begin{eqnarray}\label{lag}
\mathcal{L}=-j^\mu a_\mu+\frac{1}{2}\rho\mathbf{v}^2-V
\end{eqnarray}
where $j^\mu=(\rho,\rho\mathbf{v})$ and
\begin{equation}\label{eqn:Clebsch}
a_\mu=-\left(\partial_\mu\theta+\alpha\partial_\mu\beta\right).
\end{equation}
%Also $V$ is a $\rho,\dot{\rho}$ dependent potential which will give rise to the force. The total energy or the Hamiltonian is given by 
%\begin{eqnarray}\label{ham}
%H=\int d^3\mathbf{x}\left[\frac{1}{2}\rho\mathbf{v}^2+V(\rho,\dot{\rho})\right]
%\end{eqnarray}
%The pressure of the fluid is $P=\rho\frac{\partial V}{\partial\rho}-V$. 
Here $\theta$, $\alpha$ and $\beta$ are single valued non-singular functions which vanish at infinity. In this parametrization $\nabla\theta$ is the irrotational part of the fluid velocity field while $\nabla\times\mathbf{a}=\nabla\alpha\times\nabla\beta$ is its vorticity. The above Lagrangian (\ref{lag}) can also be written as 
\begin{eqnarray}\label{hamaction}
S=\int\left(\rho\dot{\theta}+\rho\alpha\dot{\beta}-\left[\frac{1}{2}\rho\mathbf{v}^2+V\right]\right)d^3x.
\end{eqnarray}
where $\mathbf{v}=\nabla\theta+\alpha\nabla\beta$. This type of parametrization is called Clebsch Parametrization.

%In general in fluid dynamics the Poisson Bracket of two variables (functions of $\rho$ and $v_i$) is given by 
%\begin{eqnarray*}
%[F,G]=\int\left[\frac{\delta F}{\delta \rho}\partial_i\left(\frac{\delta G}{\delta v_i}\right)-\frac{\delta G}{\delta \rho}\partial_i\left(\frac{\delta F}{\delta v_i}\right)-\frac{\omega_{ij}}{\rho}\frac{\delta F}{\delta v_i}\frac{\delta G}{\delta v_j}\right]
%\end{eqnarray*}
%Using them, we get Poisson's Bracket between $\rho$, $v_i$'s and $j$'s with $\omega_{ij}=\partial_iv_j(\mathbf{r})-\partial_jv_i(\mathbf{r})$. 

%\textcolor{red}{The Hamiltonian in (\ref{ham}) we can define \emph{helicity} defined as 
%\begin{eqnarray*}
%C=\frac{1}{8\pi}\int\epsilon^{ijk}v_i\partial_j v_k
%\end{eqnarray*}
%But this helicity commutes with all variables like $F$. So, it becomes necessary to fix a value of $C$ and then describe the dynamics of $H$. This gives the above parametrization of $\mathbf{v}$ and we can then also write down an action for fluid dynamics. It is to be noted that it's possible to give another fluid description by going to the comoving frame of the fluid and treating the particle levels continuously.}

The Clebsch parametrization can be achieved in a group theoretical language\footnote{Any group can work equally well in principle. The fact that $SU(1,1)$ works in this case, can be thought of as due to the fact that it has right number of parameters. We thank A. Polychronakos for a discussion on this point.}
using $SU(1,1)$ or $SU(2)${}\footnote{If we quantize vorticity ($\nabla\alpha\times\nabla\beta$), then we need to work with $SU(2)$.}. With an element of the group parametrized by
\begin{eqnarray*}
g=\frac{1}{\sqrt{1-\tilde{u}u}}\begin{bmatrix}1&u\\\tilde{u}&1\end{bmatrix}\begin{bmatrix}e^{i\theta/2}&0\\0&e^{-i\theta/2}\end{bmatrix}
\end{eqnarray*}
where $\tilde{u}=\bar{u}$ if $g\in\mathrm{SU}(1,1)$ and $\tilde{u}=-\bar{u}$ if $g\in\mathrm{SU}(2)$, the Clebsch parametrization is produced as
\begin{equation*}\label{eqn:ids}
-i\tr(\sigma_3g^{-1}dg)=d\theta+\alpha d\beta\quad 
\end{equation*}
where
\[
\alpha=\frac{\tilde{u}u}{1-\tilde{u}u},\quad\beta=-i\ln\left(\frac{u}{\tilde{u}}\right)
\]
$\theta$ describes the compact direction of the $U(1)$ subgroup generated by $\frac{1}{2}\sigma_3$ and $\alpha$ and $\beta$ parametrize the $SU(1,1)/U(1)$ or $SU(2)/U(1)$ coset. The action (\ref{hamaction}) can now be written as 
\begin{eqnarray*}\label{graction}
S=-i\int j^\mu\mbox{Tr}\left(\sigma_3g^{-1}\partial_\mu g\right)-\int\left[\frac{j_ij^i}{2\rho}+V\right]
\end{eqnarray*}
%Here $j^\mu=(j^0,j^i)=(\rho,j^i)$. Substituting for $j^i$ from its equation of motion, we get the action (\ref{hamaction}).
The relativistic version of such an action is given by
\begin{align*}\label{relgen}
S&=-i\int\left[ j^\mu\mbox{Tr}\left(\sigma_3g^{-1}\partial_\mu g\right)+F(n)\right]
\end{align*}
with $j^\mu=nu^\mu$ for a unit flow vector $u^\mu$, so that $j^\mu j_\mu=n^2$, where $n$ is number density, and $F(n)$ encodes the specific dynamics
(equation of state). The four velocity $u_{\mu}$ is then proportional to $a_\mu$. Note here that for free theory (no potential $V$) which describes dust, $F(n)=n$. 

%The function $F(n)$ has the interpretation of Enthalpy. This can be easily seen by calculating the stress tensor $T^{\mu\nu}$ and noting its structural similarity with the standard dust/fluid relativistic stress tensor. We find the energy density is $nF^\prime$ and pressure as $(nF^\prime-F)$. For calculational details see \cite{Bistrovic:2002jx}. 

%To end this subsection, we note that $\theta$, which parametrizes the $U(1)$ direction is compact and hence might be puzzling as from Clebsch parametrization point of view there's no need of any compact dimension. This becomes clearer if we calculate the Poisson's bracket between $\rho(f)$ ($f(x)$ is some function of $x$) and $g(x)$ (note that to calculate the Poisson's bracket, all we need is a symplectic structure which can be obtained if action is known (see e.g Nair's book)). We get 
%\begin{eqnarray*}
%[\rho(f),g(x)]=-ig(x)\frac{\sigma_3}{2}f(x)
%\end{eqnarray*}
%So in the quantum theory, for $A=exp[-2\pi i\int\rho]$, using the above Poisson's bracket result, we get $A^\dagger gA=ge^{i\pi\sigma_3}=-g$. Now from the action all the observables depend on even powers of $g$ and hence they are invariant under the action of $A$. Hence we can set $A=1$ which yields $\int\rho=N$ for some integer $N$. The existence of the compact direction can be thought of as coming from the required quantization of $\int\rho$. This also means that the fluid is made up of particles with $\rho$ being the particle density (meaning integer number of particles). 

%%%%%%%%%%%%%%%%%%%%%%%%%%%%%%%
\subsection{Non-Abelian generalization}
%%%%%%%%%%%%%%%%%%%%%%%%%%%%%%%

The generalization of the entire formalism to non-abelian case (for simplicity described in terms of $SU(2)$) at \emph{one particle level} is described by the Wong equation with action \cite{Wong:1970fu},\cite{Balachandran:1977ub}
\begin{equation*}\label{1particle}
S=\int\left[\frac{1}{2}m\dot{\mathbf{x}}^2+A^a_iQ^a\dot{x}_i-iw\mbox{Tr}(\sigma_3g^{-1}\dot{g})\right]
\end{equation*}
where $a$ indexifies the number of generators of the gauge group and $i$ the number of components of a vector. Here $Q^a=\mbox{Tr}(g\sigma_3g^{-1}t^a)$, $t^a=\frac{1}{2}\sigma^a$. For many particles the generalization is easy. With $\lambda$ indexing the particles we have 
\begin{eqnarray*}
S=-i\int dt\sum_{\lambda}w_\lambda\mbox{Tr}(\sigma_3{g_\lambda}^{-1}\dot{g_\lambda})
\end{eqnarray*}
with different $g_\lambda$'s for different particles which is in line with its continuous generalization of $g(x)$. In fact to generalize it to the continuous case, we have, $\lambda\rightarrow x$, $\sum_{\lambda}\to \int d^3\mathbf{x}/v$, $w_\lambda/v\to \rho(x)$. We then get
\begin{eqnarray*}
S=-i\int d^4x\rho\mbox{Tr}(\sigma_3g^{-1}\dot{g}).
\end{eqnarray*}
%Under a $U(1)$ transformation of the action ($g\to g \exp(i\sigma_3\phi/2)$), $\Delta S=w\Delta\phi$ and hence for a closed path of $\phi$ in $SU(2)$, the singlevaluedness of $e^{iS}$ gives quantization of $w$. 
Taking this last term as the term responsible for the symplectic structure, we can write an action
%\footnote{See section 5.3 of \cite{Jackiw:2004nm} for the construction of the action in the single particle level. This may be taken as another approach towards constructing a non-abelian action other than starting from Wong's equation.}
\begin{equation*}\label{nonabaction}
S=-i\int d^4x j^\mu\mbox{Tr}(\sigma_3g^{-1}D_\mu g)-\int F(n)+S_{YM}
\end{equation*}
where $j^{\mu}$ is the current due to the charged particle density in the fluid as described before and is again given by $j^\mu=(\rho,\mathbf{v}\rho)=nu^\mu$ with $u^2=1$.

%This is in fact a specific case of a non-abelian Lagrangian
%\begin{equation}
%\mathcal{L}=-2i\sum_{(i)}j_{(i)}^\mu\mbox{Tr}(K_{(i)}g^{-1}D_\mu g)-F(n)+\mathcal{L}_{YM}
%\end{equation}
%The current which couples to the non-abelian gauge charge $A^a_\mu$ is on the other hand given by $J^{a\mu}=\mbox{Tr}(\sigma_3g^{-1}t^ag)j^\mu=Q^aj^\mu$ ($Q^a=2\mbox{Tr}(gKg^{-1}t^a)$. Hence writing $Q=Q^at^a$, we have $Q=2\mbox{Tr}(gKg^{-1}t^a)t^a=gKg^{-1}$.\footnote{To see this note that $\mbox{Tr}(QT^a)=\frac{Q^a}{2}$ and hence $Q^a=2\mbox{Tr}(gKg^{-1}t^a)$) and is in the Eckart form (of course here $D_\mu g=\partial_\mu g+A_\mu g$, $A_\mu=-it^aA^a_\mu$, $t^a=\sigma^a/2$).} The action (\ref{nonabaction}) gives rise to proper magnetohydrodynamics and proper commutation relation between charge densities.
%
%We can also pick a specific $SU(2)$ transformation which makes charge density $\rho$ diagonal. The charge density matrix $\rho$ transforms as a matrix in the fundamental representation of $SU(2)$ as $\rho\to h^{-1}\rho h$, $h\in SU(2)$, $\rho=\rho^at^a$. Hence the $SU(2)$ transformation $g$ will be like $\rho=g\rho_{\mbox{diag}}g^{-1}$, so that $\rho^a=n\mbox{Tr}(g\sigma_3g^{-1}t^a)=nQ^a$. Hence we see that $g(\mathbf{x},t)$ are related to charge densities. Also the eigenvalues of $\rho$ are gauge invariant and represented by $n$.
%
%MAY BE WE CAN SKIP A FEW THINGS HERE IN THE LAST TWO PARAGRAPHS??

For a general gauge group $G$, the action is given by 
\begin{eqnarray*}
S=-i\int\sum_{(i)}j_{(i)}^\mu\mbox{Tr}(K_{(i)}g^{-1}D_\mu g)-\int F(n_1,n_2,\dots)+S_{YM}(A)
\end{eqnarray*}
Here $K_{(i)}$ are diagonal generators of $G$ and $j_{(i)}^\mu j_{(i)\mu}=n_{(i)}^{2}$ with $i$ indexing the rank of the group $G$.

%%%%%%%%%%%%%%%%%%%%%%%%%%%%%%%%%%%%
\section{Spinning Fluids Using the de Sitter Group}
\label{sec:dS_Spinning_fluids}
%%%%%%%%%%%%%%%%%%%%%%%%%%%%%%%%%%%%

From the models reviewed in section~[\ref{sec:intro}] it would be natural to write down a similar action describing the dynamics of fluids in terms of the Poincar\'e group where the information about the spin would be contained in the Lorentz subgroup and the information about the mass current would be contained in the translational part. As we will see in section~[\ref{sec:alternative_approaches}] this creates internal inconsistency solved by the introduction of a constraint that reduces the degrees of freedom of the dynamics. Moreover the theory of Gravity cannot be written in terms of a gauge theory of the Poincar\'e group directly. Only starting from the de~Sitter group and then using a reduction limit we can reproduce gravity as a gauge theory. With this idea in mind we propose a generalization of the spin fluid action introduced in \cite{Nair:2011mk} with the following action to describe the full dynamics of relativistic fluids with spin for a torsion free background\footnote{For a discussion about torsion free models, see \cite{Nair:2011mk}.}:
\begin{equation}\label{eqn:dSaction}
S=
-\int\det e\left[
ij_{(m)}^{\mu}\tr(\alpha_{(m)}T_{0}g^{-1}\nabla_{\mu}g)
+ij_{(s)}^{\mu}\tr(\alpha_{(s)}T_{23}g^{-1}\nabla_{\mu}g)
+F(n_{(m)},n_{(s)})
\right]
\end{equation}
where $g$ is a $SO(4,1)$-valued field and $\alpha_{(m)}$ and $\alpha_{(s)}$ are constants respectively for the mass and spin coupling. The choice of the Lie algebra elements of $so(4,1)$ is taken considering that we are interested in studying matter fluids ($T_{0}$ generates a time-like velocity) with spin, generated by $T_{23}$ (see appendix~\ref{appendixC} for a brief description of the generators of $so(4,1)$).

Action (\ref{eqn:dSaction}) is invariant separately under $g\to ge^{-i\gamma_{0}T^{0}}$ and $g\to ge^{-i\gamma_{23}T^{23}}$ with $\gamma_{0}$ and $\gamma_{23}$ constants; the associated conserved currents are $j_{(i)}^{\mu}$ with
\begin{equation}\label{eqn:dSDj=0}
\nabla_{\mu}j_{(i)}^{\mu}=0
\end{equation}
where $i=\{m,s\}$.

The explicit form of the current is
\begin{equation}\label{eqn:dSj_i}
j_{(i)\mu}=-i\frac{n_{(i)}}{F_{(i)}}\tr[K_{(i)}g^{-1}\nabla_{\mu}g]
\end{equation}
where $F_{(i)}=\partial F/\partial n_{(i)}$, $K_{(m)}=\alpha_{(m)}T_{0}$ and $K_{(s)}=\alpha_{(s)}T_{23}$.

The variation of the action with respect to $g$ yields
\begin{equation}\label{eqn:dSDJ=0}
\nabla_{\mu}\left(
j_{(m)}^{\mu}gK_{(m)}g^{-1}
+j_{(s)}^{\mu}gK_{(s)}g^{-1}
\right)=
0
\end{equation}
that states the conservation of the $so(4,1)$-current
\begin{equation}\label{eqn:dSJ}
J^{\mu}
=j_{(m)}^{\mu}gK_{(m)}g^{-1}
+j_{(s)}^{\mu}gK_{(s)}g^{-1}
\end{equation}
Using (\ref{eqn:dSDj=0}) we can rewrite (\ref{eqn:dSDJ=0}) as follows
\begin{equation}\label{eqn:dS[gdg,J]}
j_{(m)}^{\mu}[g^{-1}\nabla_{\mu}g,K_{(m)}]
+j_{(s)}^{\mu}[g^{-1}\nabla_{\mu}g,K_{(s)}]
=0
\end{equation}
where we used (\ref{eqn:dSDj=0}).

The components (refer to appendix~\ref{appendixC} for the notation for the normalized trace) of the $gK_{(i)}g^{-1}$
\[
Q_{(i)}^{AB}=-2\tr[gK_{(i)}g^{-1}T^{AB}]
\]
are the non-abelian charges of the theory (\ref{eqn:dSJ}). These charges are the coefficients of $\delta w_{\mu ab}$ in the variation of the action with respect to the frame fields. In this sense they are interpreted as spin charges. In particular $Q_{(s)}^{ab}$ is the spin charge of the fluid, been generated by the spin degrees of freedom, and $Q_{(m)}^{ab}$ is related to the vorticity of the fluid, being related to the mass current of the fluid. We will discuss more about the relation between $Q_{(m)}^{ab}$ and the vorticity in a short while.

Equations (\ref{eqn:dSj_i}) can be used to write a Mathisson-Papapetrou equation for fluids. We apply the covariant derivative and then we anti-symmetrize the two indices and contract one index with the current to get the following expression:
%
%\[
%\nabla_{\nu}\left(
%\frac{F^{(i)}}{n_{(i)}}j_{i\mu}
%\right)
%=i\tr[gK_{(i)}g^{-1}\nabla_{\nu}gg^{-1}\nabla_{\mu}gg^{-1}]
%-i\tr[gK_{(i)}g^{-1}\nabla_{\nu}\nabla_{\mu}gg^{-1}]
%\]
%
%\[
%\nabla_{[\nu}\left(
%\frac{F^{(i)}}{n_{(i)}}j_{i\mu]}
%\right)
%=i\tr[gK_{(i)}g^{-1}\nabla_{[\nu}gg^{-1}\nabla_{\mu]}gg^{-1}]
%-\frac{i}{2}\tr[gK_{(i)}g^{-1}[\nabla_{\nu},\nabla_{\mu}]gg^{-1}]
%\]\[
%=-\frac{i}{2}\tr[\nabla_{\nu}\left(gK_{(i)}g^{-1}\right)\nabla_{\mu}gg^{-1}]
%-\frac{i}{2}\tr[gK_{(i)}g^{-1}R_{\nu\mu}]
%\]
%
%Then we can rewrite the equation for the currents as
\begin{equation}\label{eqn:dSMP_i}
j_{(i)}^{\nu}\nabla_{[\nu}\left(
\frac{F_{(i)}}{n_{(i)}}j_{(i)\mu]}
\right)
=-\frac{i}{2}\tr\left[j_{(i)}^{\nu}\nabla_{\nu}\left(gK_{(i)}g^{-1}\right)\nabla_{\mu}gg^{-1}\right]
-\frac{i}{2}\tr\left[gK_{(i)}g^{-1}R_{\nu\mu}\right]j_{(i)}^{\nu}
\end{equation}

Adding the two equations in (\ref{eqn:dSMP_i}), and using (\ref{eqn:dSDJ=0}), we have the generalized Mathisson-Papapetrou equation for fluids
\begin{eqnarray}
&&j_{(m)}^{\nu}\nabla_{[\nu}\left(
\frac{F_{(m)}}{n_{(m)}}j_{(m)\mu]}
\right)
+j_{(s)}^{\nu}\nabla_{[\nu}\left(
\frac{F_{(s)}}{n_{(s)}}j_{(s)\mu]}
\right)=\nonumber\\
&&=-\frac{i}{2}\tr[gK_{(m)}g^{-1}R_{\nu\mu}]j_{(m)}^{\nu}
-\frac{i}{2}\tr[gK_{(s)}g^{-1}R_{\nu\mu}]j_{(s)}^{\nu}.\label{eqn:dSMP}
\end{eqnarray}
Notice that only the $Q^{ab}_{(i)}$ components of the charge contributes to the right hand side.

Equation (\ref{eqn:dSMP}) can also be derived from the conservation of the energy-momentum tensor of the fluid given below:
\begin{eqnarray}
T^{\mu\nu}\label{eqn:dST}
\!\!\!&=&\!\!\!
-\left[F_{(m)}n_{(m)}+F_{(s)}n_{(s)}-F(n_{(m)},n_{(s)})\right]g^{\mu\nu}
+\frac{F_{(m)}}{n_{(m)}}j_{(m)}^{\mu}j_{(m)}^{\nu}
+\frac{F_{(s)}}{n_{(s)}}j_{(s)}^{\mu}j_{(s)}^{\nu}
\nonumber\\
&&\;\;\;+2\nabla_{\gamma}\left(
j_{(m)}^{\mu}Q_{(m)}^{\gamma\nu}+j_{(m)}^{\nu}Q_{(m)}^{\gamma\mu}
\right)
+2\nabla_{\gamma}\left(
j_{(s)}^{\mu}Q_{(s)}^{\gamma\nu}+j_{(s)}^{\nu}Q_{(s)}^{\gamma\mu}
\right)
\end{eqnarray}
Setting $Q^{ab}_{(i)}=0$ (and $\alpha_{(s)}=0$ if we want to discuss only about the mass current), the energy-momentum tensor (\ref{eqn:dST}) reduces to the one for an ideal fluid with
\begin{eqnarray}
p &=& F_{(m)}n_{(m)}+F_{(s)}n_{(s)}-F(n_{(m)},n_{(s)})\nonumber\\
\rho &=& F\label{eqn:dSIdeal}\\
p+\rho &\equiv & \mu_{(m)}n_{(m)}+\mu_{(s)}n_{(s)} = F_{(m)}n_{(m)}+F_{(s)}n_{(s)}\nonumber
\end{eqnarray}
where the $\mu_{(i)}=F_{(i)}=\partial U/\partial N_{(i)}$ are the chemical potentials with $U=\rho V$ the internal energy and $N_{(i)}$ the particle number. Notice that the Mathisson-Papapetrou equation (\ref{eqn:dSMP}) is also easily derived from the conservation of the energy-momentum tensor
\[
\nabla_{\mu}T^{\mu\nu}
=2j_{(m)\mu}\nabla^{[\mu}\left(\frac{F_{(m)}}{n_{(m)}}j_{(m)}^{\nu]}\right)
+2j_{(s)\mu}\nabla^{[\mu}\left(\frac{F_{(s)}}{n_{(s)}}j_{(s)}^{\nu]}\right)
+2R^{\nu}_{\phantom{-}\rho\mu\gamma}\left(
j_{(m)}^{\rho}Q_{(m)}^{\gamma\mu}
+j_{(s)}^{\rho}Q_{(s)}^{\gamma\mu}
\right)
=0
\]
where we used the equality
\[
R^{\nu}_{\phantom{-}\rho\mu\gamma}j^{\mu}Q^{\gamma\rho}
+\frac{1}{2}R^{\nu}_{\phantom{-}\rho\gamma\mu}j^{\rho}Q^{\gamma\mu}
=0
\]
which is the consequence of the anti-symmetry of $Q_{(i)}^{\alpha\beta}$ and $R^{\nu}_{\phantom{-}[\rho\mu\gamma]}=0$.

In section~\ref{sec:reduction_limit} we will perform the reduction limit of these equations. This procedure will reproduce the equation of motion for the fluid. The study of the effect of higher order corrections and the consequences of the possible extension of the model to the full de Sitter group are possible directions to take for future research.

%%%%%%%%%%%%%%%%%%%%%%%%%%%%%%%%%%%%%%%%%%%%%%%%%%%%%%
\subsection{Reduction Limit}\label{sec:reduction_limit}
%%%%%%%%%%%%%%%%%%%%%%%%%%%%%%%%%%%%%%%%%%%%%%%%%%%%%%

For the reduction limit we are going to consider the following parametrization
\begin{equation}\label{eqn:g=Lh}
g=\Lambda e^{-i\frac{\theta^{a}T_{a}}{R}}=\Lambda h
\qquad\textrm{with }\Lambda\in\mathrm{SO}(3,1)
\end{equation}
that describes a subset of the de Sitter group. The Lorentz Group is reproduced in the $R\to\infty$ limit. Also noticing that
\begin{eqnarray}
gK_{(m)}g^{-1} &\sim&
\alpha_{(m)}\left(
\Lambda T^{0}\Lambda^{-1}
-\frac{i}{2R}\theta_{a}\Lambda T^{a0}\Lambda^{-1}
\right)
+\mathcal{O}\left(\frac{1}{R^{2}}\right)\label{eqn:dSR_Q_m}\\
gK_{(s)}g^{-1} &\sim&
\alpha_{(s)}\Lambda T^{23}\Lambda^{-1}
+\mathcal{O}\left(\frac{1}{R}\right)\label{eqn:dSR_Q_s}
\end{eqnarray}
we re-establish the right proportion replacing $\alpha_{(s)}$ with $\alpha_{(s)}/R$.

The currents (\ref{eqn:dSj_i}) reduce to
\begin{eqnarray}
j_{(m)\mu}
&=&\frac{n_{(m)}}{4F_{(m)}}\frac{\alpha_{(m)}}{R}\partial_{\mu}\theta_{0}
-\frac{n_{(m)}}{F_{(m)}}\tr\left[
\frac{\theta^a}{2R}\alpha_{(m)}T_{a0}\Lambda^{-1}\nabla_{\mu}\Lambda
\right]
+\mathcal{O}\left(\frac{1}{R^3}\right)\label{eqn:rdj_m}\\
j_{(s)\mu}
&=&-\frac{in_{(s)}}{F_{(s)}}\tr\left[
\frac{\alpha_{(s)}}{R}T_{23}\Lambda^{-1}\nabla_{\mu}\Lambda
\right]
+\mathcal{O}\left(\frac{1}{R^2}\right)\label{eqn:rdj_s}
\end{eqnarray}
So, we see that both the $j_{(i)}^{\mu}$'s are of the order $\mathcal{O}\left(\frac{1}{R}\right)$. Hence we get
\[
n_{(i)}
=\sqrt{j^{\mu}_{(i)}j_{(i)\mu}}
\sim\mathcal{O}\left(\frac{1}{R}\right)
\]
implying that also $F\sim\mathcal{O}\left(\frac{1}{R^{2}}\right)$ and $F_{(i)}\sim\mathcal{O}\left(\frac{1}{R}\right)$. On the other hand, the $so(4,1)$-valued current (\ref{eqn:dSJ}) reduces to
\begin{align*}
J^\mu&=j^\mu_{(m)}\alpha_{(m)}\Lambda\left[
T^{0}
-i\frac{\theta_{a}}{2R}T^{a0}
\right]\Lambda^{-1}+j^\mu_{(s)}\frac{\alpha_{(s)}}{R}\Lambda
T^{23}
\Lambda^{-1}+\mathcal{O}\left(\frac{1}{R^3}\right)\nonumber\\
&=j^\mu_{(m)}\alpha_{(m)}v_aT^{a}+\frac{1}{R}\left[-i\frac{\theta_{a}}{2}j_{(m)}^{\mu}\alpha_{(m)} \Lambda T^{a0}\Lambda^{-1}
+j_{(s)}^{\mu}\alpha_{(s)}
\Lambda T^{23}\Lambda^{-1}\right]+\mathcal{O}\left(\frac{1}{R^3}\right)
\end{align*}
where $v^{a}=\Lambda^{a}_{\;\;0}$.

To evaluate the reduction limit of (\ref{eqn:dSDJ=0}) we replace $g$ with the parametrization (\ref{eqn:g=Lh}) and substitute it in (\ref{eqn:dS[gdg,J]}) obtaining
\[
[
\Lambda^{-1}\nabla_{\mu}\Lambda,\alpha_{(m)}j^{\mu}_{(m)}T^{0}
]
+\frac{1}{R}\left[
i\theta_{a}[T^{a},\Lambda^{-1}\nabla_{\mu}\Lambda]
-i\partial_{\mu}\theta_{a}T^{a},
j^{\mu}_{(m)}\alpha_{(m)}T^{0}
\right]
+\frac{1}{R}[\Lambda^{-1}\nabla_{\mu}\Lambda,j^{\mu}_{(s)}\alpha_{(s)}T_{23}]
=0,
\]
where we kept terms until the $\mathcal{O}(1/R^{2})$. Evaluating the trace we get the following equations:
\begin{itemize}

\item the $T^{a}$ ($a\neq 0$) component
\begin{equation}\label{eqn:dSDJ^a}
\alpha_{(m)}j^{\mu}_{(m)}(\Lambda^{-1}\nabla_{\mu}\Lambda)^{0}_{\;\;b}
=0;
\end{equation}

\item the $T^{10}$ component
\begin{equation}\label{eqn:dSDJ^10}
\frac{i}{2}\alpha_{(m)}j^{\mu}_{(m)}\left[
\theta^{b}(\Lambda^{-1}\nabla_{\mu}\Lambda)_{b}^{\;\;1}-\partial_{\mu}\theta^{1}
\right]
-2\alpha_{(s)}j^{\mu}_{(s)}(\Lambda^{-1}\nabla_{\mu}\Lambda)^{0[2}\eta^{3]1}
=0;
\end{equation}

\item the $T^{20}$ component
\begin{equation}\label{eqn:dSDJ^20}
\frac{i}{2}\alpha_{(m)}j^{\mu}_{(m)}\left[
\theta^{b}(\Lambda^{-1}\nabla_{\mu}\Lambda)_{b}^{\;\;2}-\partial_{\mu}\theta^{2}
\right]
-2\alpha_{(s)}j^{\mu}_{(s)}(\Lambda^{-1}\nabla_{\mu}\Lambda)^{0[2}\eta^{3]2}
=0;
\end{equation}

\item the $T^{30}$ component
\begin{equation}\label{eqn:dSDJ^30}
\frac{i}{2}\alpha_{(m)}j^{\mu}_{(m)}\left[
\theta^{b}(\Lambda^{-1}\nabla_{\mu}\Lambda)_{b}^{\;\;3}-\partial_{\mu}\theta^{3}
\right]
-2\alpha_{(s)}j^{\mu}_{(s)}(\Lambda^{-1}\nabla_{\mu}\Lambda)^{0[2}\eta^{3]3}
=0;
\end{equation}

\item the $T^{12}$ component
\begin{equation}\label{eqn:dSDJ^12}
2\alpha_{(s)}j^{\mu}_{(s)}(\Lambda^{-1}\nabla_{\mu}\Lambda)^{13}
=0;
\end{equation}

\item the $T^{13}$ component
\begin{equation}\label{eqn:dSDJ^13}
-2\alpha_{(s)}j^{\mu}_{(s)}(\Lambda^{-1}\nabla_{\mu}\Lambda)^{12}
=0.
\end{equation}

\end{itemize}

Notice that the $T^{23}$ and the $T^{0}$ component are zero at all orders having
\[
\left[
\alpha_{(m)}j_{(m)}^{\mu}T^{0}+\frac{\alpha_{(s)}}{R}j^{\mu}_{(s)}T^{23},T^{0}
\right]=0 \quad\mbox{and}\quad
\left[
\alpha_{(m)}j_{(m)}^{\mu}T^{0}+\frac{\alpha_{(s)}}{R}j^{\mu}_{(s)}T^{23},T^{23}
\right]=0
\]
Multiplying equations~(\ref{eqn:dSDJ^10}), (\ref{eqn:dSDJ^20}), and (\ref{eqn:dSDJ^30}) respectively by $\theta_{1}$, $\theta_{2}$, and $\theta_{3}$, and adding them, we find
\begin{equation}
-\frac{i}{2}\alpha_{(m)}j^{\mu}_{(m)}\sum_{a=1}^{3}\partial_{\mu}\theta^{a}\theta_{a}
-2\alpha_{(s)}j^{\mu}_{(s)}(\Lambda^{-1}\nabla_{\mu}\Lambda)^{0[2}\theta^{3]}
=0.
\end{equation}
Moreover equation~(\ref{eqn:dSDJ^a}) can be rewritten as follows
\[
\alpha_{(m)}j_{(m)}^{\mu}(\Lambda^{-1})^{b}_{\;\;c}e^{c}_{\alpha}[
\partial_{\mu}(e^{\alpha}_{a}\Lambda^{a}_{\;\;0})
+\Gamma^{\alpha}_{\mu\beta}e^{\beta}_{a}\Lambda^{a}_{\;\;0}
]=0;
\]
the above relation suggests that
\begin{equation}\label{eqn:dSv}
v^{\alpha}=e^{\alpha}_{a}\Lambda^{a}_{\;\;0}
\end{equation}
is related to the fluid velocity field and for the case of dust with zero vorticity satisfying the relativistic Euler equation (\ref{eqn:EulerDust}), it is precisely the fluid velocity (as discussed in section~\ref{sec:dust}).

From the definition we have seen that the charges $Q_{(m)}^{ab}$ are proportional to the spatial part of $\theta_{a}$ which appears in several of the above equations. Moreover, defining the vorticity as in \cite{Andersson:2006nr}
\[
\omega_{\nu\mu}=\nabla_{[\nu}\left(\frac{\mu_{(m)}}{n_{(m)}}j_{(m)\mu]}\right)
\]
we find\footnote{Notice that this recovers the definition in terms of the Clebsch parametrization.}
\[
\omega_{\nu\mu}
=\nabla_{[\nu}\left(
\frac{F_{(m)}}{n_{(m)}}j_{(m)\mu]}
\right)
=-\frac{i}{2}\tr[\nabla_{\nu}\left(gK_{(m)}g^{-1}\right)\nabla_{\mu}gg^{-1}]
-\frac{i}{2}\tr[gK_{(m)}g^{-1}R_{\nu\mu}]
=0
\]
if $Q_{(m)}^{ab}=0$. Hence we are going to consider the $Q_{(m)}^{ab}$ along with the spatial part of $\theta^{a}$ to be the ``charges'' of the vorticity of the fluid. In section~\ref{sec:Godel} we will study a case with non-zero $\theta^{i}$'s and vorticity.

Finally the Mathisson-Papapetrou equation for spinning fluids in the reduction limit reduces to
\begin{eqnarray}\label{eq:MP2}
&&j_{(m)}^{\nu}\nabla_{[\nu}\left(\frac{F_{(m)}}{n_{(m)}}j_{(m)\mu]}\right)
+j_{(s)}^{\nu}\nabla_{[\nu}\left(\frac{F_{(s)}}{n_{(s)}}j_{(s)\mu]}\right)=
\nonumber\\&&=-\frac{i}{2}\tr\left[
\Lambda\left(-i\frac{\theta^{a}}{2R}j^{\nu}_{(m)}\alpha_{(m)}T_{a0}
+j^\nu_{(s)}\frac{\alpha_{(s)}}{R}T_{23}
\right)\Lambda^{-1}R_{\nu\mu}\right]
\end{eqnarray}
Notice that the equation is of order $\mathcal{O}\left(\frac{1}{R^2}\right)$, as the energy-momentum tensor. This implies that, before taking the limit we also need to require the gravitational side to be of order $\mathcal{O}\left(\frac{1}{R^2}\right)$.

We are going to end this section by checking that, in the reduction limit, the two Casimirs of $SO(4,1)$ reproduce the correct conserved quantities. The first Casimir is given by
\[
C_{2}=
-\eta_{\mu\nu}\tr[J^{\mu}J^{\nu}]
=\frac{n_{(m)}^{2}\alpha_{(m)}^{2}+n_{(s)}^{2}\alpha_{(s)}^{2}/R^{2}}{4}
\to\frac{n_{(m)}^{2}\alpha_{(m)}^{2}}{4}
\]
and is related to the conservation of mass for a particle, $n_{(m)}$ being the number density times the mass. The second Casimir is given by
\[
g^{\mu\mu'}J^{ab}_{\mu}J^{c}_{\nu}\varepsilon_{abcd}g^{\nu\nu'}J_{\mu'a'b'}J_{\nu'c'}\varepsilon^{a'b'c'd}
\to \frac{\alpha_{(m)}^{2}\alpha_{(s)}^{2}n_{(m)}^{2}n_{(s)}^{2}}{R^{2}}Q_{(s)}^{ab}v^{c}\varepsilon_{abcd}Q_{(s)a'b'}v_{c'}\varepsilon^{a'b'c'd}
\]
that is proportional to the module squared of the Pauli-Lubanski vector, confirming our interpretation of $Q_{(s)}$ as the spin charge of the fluid.

\subsection{Dust limit}
\label{sec:dust}

From the energy-momentum tensor (\ref{eqn:dST}), once we neglect the terms containing the vorticity and the spin-density, it is easy to recognize the density and the pressure of an ideal fluid.
\[
\rho+p=n_{(m)}F_{(m)}+n_{(s)}F_{(s)} \qquad
p=n_{(m)}F_{(m)}+n_{(s)}F_{(s)}-F,
\]
that is,
\begin{equation}\label{eqn:rho,P}
\rho=F \qquad
p=n_{(m)}F_{(m)}+n_{(s)}F_{(s)}-F.
\end{equation}

For spinless dust we can set
\[
F(n_{(m)},n_{(s)})=n_{(m)} \qquad\textrm{and}\qquad
\alpha_{(s)}=0,
\]
that is,
\[
j_{(m)\mu}=-in_{(m)}\tr[K_{(m)}g^{-1}\nabla_{\mu}g] \qquad\textrm{and}\qquad
j_{(s)\mu}=0.
\]
The Mathisson-Papapetrou equation (\ref{eq:MP2}) then reduces to
\[
j_{(m)}^{\nu}\nabla_{\nu}\left(
\frac{1}{n_{(m)}}j_{(m)\mu}
\right)
=-i\tr\left[\Lambda\left(-i\frac{\theta^{a}}{2R}j^\mu_m\alpha_{(m)} T_{a0}
\right)\Lambda^{-1}R_{\nu\mu}\right]
\]
This recovers the relativistic Euler equation for a fluid of dust with zero vorticity
\begin{equation}\label{eqn:EulerDust}
\nabla_{\mu}j_{(m)}^{\mu}=0 \qquad
\nabla_{[\nu}\left(
\frac{1}{n_{(m)}}j_{(m)\mu]}
\right)
=0
\end{equation}
in the case $\theta^a=(\theta^{0},0,0,0)$. Therefore from
\[
j^\mu_{(m)}
=\frac{n_{(m)}}{4}\alpha_{(m)}\partial^{\mu}\theta_{0}
=n_{(m)}u^{\mu}
\]
we can recover $\theta_{0}$ once $u^{\mu}$ is known. Notice that such an expression is exactly the same as that of a pressureless fluid with zero vorticity in the Clebsch parametrization (\ref{eqn:Clebsch}).

The remaining part of the group-valued field $\Lambda$ can be obtained from (\ref{eqn:dSDJ^a}) and (\ref{eqn:EulerDust}), which yields (\ref{eqn:dSv})
\[
\frac{1}{n_{(m)}}j_{(m)}^{\mu}=e^{\mu}_{a}\Lambda^{a}_{\;\;0}
\]
This tells us that $v^{a}e_{a}^{\mu}=\Lambda^{a}_{\;\;0}e_{a}^{\mu}$ is nothing else than the fluid velocity $u^{\mu}$ with respect to the local frame described by the frame field. Notice that the above statement is always true in the case of dust with zero vorticity.
\newline
\newline
Let us now consider the case of dust with spin. For dust with spin we need to set
\[
F(n_{(m)},n_{(s)})=n_{(m)}+n_{(s)}.
\]
The currents will be
\begin{align}
j_{(m)\mu}
&=\frac{n_{(m)}}{4}\frac{\alpha_{(m)}}{R}\left[
\partial_{\mu}\theta_{0}
+\theta^{a}(\Lambda^{-1}\nabla_{\mu}\Lambda)_{a0}
\right]\\
j_{(s)\mu}
&=\frac{in_{(s)}}{2}\frac{\alpha_{(s)}}{R}(\Lambda^{-1}\nabla_{\mu}\Lambda)_{23}.
\end{align}
The Mathisson-Papapetrou equation (\ref{eq:MP2}) then reduces to
\[
j_{(m)}^{\nu}\nabla_{\nu}\left(\frac{1}{n_{(m)}}j_{(m)\mu}\right)+j_{(s)}^{\nu}\nabla_{\nu}\left(\frac{1}{n_{(s)}}j_{(s)\mu}\right)=-i\tr\left[\Lambda\left(-i\frac{\theta^{a}}{2R}j^\mu_{(m)}\alpha_{(m)} T_{a0}
+j^\mu_{(s)}\frac{\alpha_{(s)}}{R}
T_{23}\right)\Lambda^{-1}R_{\nu\mu}\right]
\]
that recover the expected result (the Mathisson-Papapetrou equation for many non-interacting particles) in the zero-vorticity case, $\theta^a=(\theta^0,0,0,0)$,
\begin{eqnarray}\label{oldresult}
j_{(m)}^{\nu}\nabla_{\nu}\left(\frac{1}{n_{(m)}}j_{(m)\mu}\right)+j_{(s)}^{\nu}\nabla_{\nu}\left(\frac{1}{n_{(s)}}j_{(s)\mu}\right)=-i\tr\left(
j^\mu_{(s)}\frac{\alpha_{(s)}}{R}
\Lambda T_{23}\Lambda^{-1}R_{\nu\mu}\right).
\end{eqnarray}

\subsection{Some Warm-up Examples}

In the next two subsections, we describe two simple textbook examples to show how in these cases we obtain the expected results and there exists a group valued field that parametrizes such cases. The aim of these sections is to learn the technicalities and complexities that one would encounter in realistic cases.

\subsubsection{Fluids in FRW spacetime}\label{section:FRW}

In this section we are going to solve the fluid equations in an FRW background. Because of our initial choice of the action, only the matter dominated universe can be studied. The radiation dominated Universe would require a modified action with the correct requirement of producing a light-like current along with $T_{0}\to T_{0}\pm T_{1}$.

We start with the general FRW-metric which is
\begin{equation*}
ds^{2}
=dt^{2}
-a^{2}(t)\left[
\frac{dr^{2}}{1-\kappa r^{2}}
+r^{2}(d\theta^{2}+sin^{2}\theta d\phi^{2})
\right]
\end{equation*}
The stationary currents
\begin{align*}
j_{(m)t}
&=\frac{n_{(m)}}{4F_{(m)}}\frac{\alpha_{(m)}}{R}\left[
\partial_{t}\theta_{0}
+\theta^{a}(\Lambda^{-1}\nabla_{t}\Lambda)_{a0}
\right]\\
j_{(s)t}
&=\frac{iRn_{(s)}}{2F_{(s)}}\frac{\alpha_{(s)}}{R}(\Lambda^{-1}\nabla_{t}\Lambda)_{23}
\end{align*}
depend only on the cosmological time for the case of an isotropic and homogeneous space-time under discussion. The conservation of these currents (\ref{eqn:dSDj=0}) yields
\begin{equation}\label{eqn:cosmo_Dj=0}
\partial_{t}j_{(i)}^{t}+3\frac{\dot{a}}{a}j_{(i)}^{t}=0 \qquad\Rightarrow\qquad
j_{(i)}^{t}=\beta_{(i)}a^{-3},
\end{equation}
with $\beta_{(i)}$ constant, and
\begin{eqnarray*}
j^{t}_{(m)}\left(\partial_{t}Q_{(m)}^{ti}+\frac{\dot{a}}{a}Q_{(m)}^{ti}\right)
+j^{t}_{(s)}\left(\partial_{t}Q_{(s)}^{ti}+\frac{\dot{a}}{a}Q_{(s)}^{ti}\right)
&=&0\\
j^{t}_{(m)}\left(\partial_{t}Q_{(m)}^{ij}+2\frac{\dot{a}}{a}Q_{(m)}^{ij}\right)
+j^{t}_{(s)}\left(\partial_{t}Q_{(s)}^{ij}+2\frac{\dot{a}}{a}Q_{(s)}^{ij}\right)
&=&0
\end{eqnarray*}
where the $Q_{(i)}$'s are only function of the cosmological time.

%Solving the above equations we find
%\[
%Q_{(i)}^{0i}=\gamma_{(i)}^{0i}a^{-1} \qquad
%Q_{(i)}^{ij}=\delta_{(i)}^{ij}a^{-2}
%\]
%where $\beta_{(i)},\gamma_{(i)}^{0i},\delta_{(i)}^{ij}$ are generic constants.

The components of the energy-momentum tensor (\ref{eqn:dST}) in our case are:
\begin{eqnarray}
T^{tt}
\!\!\!\!&=&\!\!\!\!
F
+4\frac{2-kr^{2}}{r(1-kr^{2})}(j_{(m)}^{t}Q_{(m)}^{rt}+j_{(s)}^{t}Q_{(s)}^{rt})
+4\cot\theta (j_{(m)}^{t}Q_{(m)}^{\theta t}+j_{(s)}^{t}Q_{(s)}^{\theta t})
\\
T^{it}
\!\!\!\!&=&\!\!\!\!
4\frac{\dot{a}}{a}(j_{(m)}^{t}Q_{(m)}^{ti}+j_{(s)}^{t}Q_{(s)}^{ti})
+\frac{2(2-kr^{2})}{r(1-kr^{2})}(j_{(m)}^{t}Q_{(m)}^{ri}+j_{(s)}^{t}Q_{(s)}^{ri})\nonumber\\
&&\!\!\!\!+2\cot\theta (j_{(m)}^{t}Q_{(m)}^{\theta i}+j_{(s)}^{t}Q_{(s)}^{\theta i})
\\
T^{ij}
\!\!\!\!&=&\!\!\!\!
-(F_{(m)}n_{(m)}+F_{(s)}n_{(s)}-F)g^{ij}
\end{eqnarray}
We can also readily write Einstein's equations as
\begin{eqnarray}
\frac{\dot{a}^{2}}{a^{2}}
+\frac{k}{a^{2}}
&\!\!\!\!=\!\!\!\!&
\!\!\frac{8\pi}{3}\!\!\left[
F
+\frac{4(2-kr^{2})}{r(1-kr^{2})}(j_{(m)}^{t}Q_{(m)}^{rt}+j_{(s)}^{t}Q_{(s)}^{rt})
+4\cot\theta (j_{(m)}^{t}Q_{(m)}^{\theta t}+j_{(s)}^{t}Q_{(s)}^{\theta t})
\right]\label{eqn:cosmoEtt}\\
2\frac{\ddot{a}}{a}
&\!\!\!\!=\!\!\!\!&
-\frac{\dot{a}^{2}}{a^{2}}-\frac{k}{a^{2}}
-8\pi(F_{(m)}n_{(m)}+F_{(s)}n_{(s)}-F)\label{eqn:cosmoEij}\\
0
&\!\!\!\!=\!\!\!\!&4\frac{\dot{a}}{a}(j_{(m)}^{t}Q_{(m)}^{ti}+j_{(s)}^{t}Q_{(s)}^{ti})
+\frac{2(2-kr^{2})}{r(1-kr^{2})}(j_{(m)}^{t}Q_{(m)}^{ri}+j_{(s)}^{t}Q_{(s)}^{ri})
\nonumber\\
&&+2\cot\theta (j_{(m)}^{t}Q_{(m)}^{\theta i}+j_{(s)}^{t}Q_{(s)}^{\theta i})\label{eqn:cosmoEti}
\end{eqnarray}

The lhs of equation (\ref{eqn:cosmoEtt}) does depend on the cosmological time $t$ only, while the rhs contains also other coordinates in the coefficient of the $Q_{(i)}$'s. Equation (\ref{eqn:cosmoEtt}) is then well posed only if all the $Q_{(i)}$'s, that are functions only of $t$, are zero or if
\[
j_{(m)}^{t}Q_{(m)}^{\mu\nu}+j_{(s)}^{t}Q_{(s)}^{\mu\nu}=0.
\]
This second option is described by a proportionality relation between $\theta^{c}\Lambda^{a}_{\;[c}\Lambda^{b}_{\;0]}$ and $\Lambda^{a}_{\;[2}\Lambda^{b}_{\;3]}$, that cannot be achieved. Therefore the only solution is to take all the $Q_{(i)}$'s to be zero
\begin{eqnarray*}
Q_{(s)}^{ab}=0 &\Rightarrow& \alpha_{(s)}=0\\
Q_{(m)}^{ab}=0 &\Rightarrow& \theta^{a}=(\theta^{0},0,0,0),
\end{eqnarray*}
that is, to consider a fluid with zero-vorticity and zero spin-density. This is not unexpected considering that the FRW-background is homogeneous and isotropic. 

From the Einstein's equations we recover the classical FRW cosmologies. In particular let us restrict to the case of dust where
\[
F=\rho_{0}n_{(m)},
\]
with $\rho_{0}$ constant. We find:
\begin{eqnarray}
3\frac{\dot{a}^{2}}{a^{2}}
+3\frac{k}{a^{2}}
&=&8\pi F\nonumber\\
\left(
2\frac{\ddot{a}}{a}+\frac{\dot{a}^{2}}{a^{2}}+\frac{k}{a^{2}}
\right)
&=&-8\pi(F_{(m)}n_{(m)}+F_{(s)}n_{(s)}-F)\nonumber
\end{eqnarray}
from which we can easily read (\ref{eqn:rho,P})
\[
\rho=F=\rho_{0}n_{(m)} \qquad\mathrm{and}\qquad
P=F_{(m)}n_{(m)}-F=0.
\]
The constant $\rho_{0}$ is just a scaling factor that can be reabsorbed by redefining $n_{(s)}$ and $j_{(m)}$ to set $\rho=n_{(m)}$. In particular in equation (\ref{eqn:cosmo_Dj=0}) we recognize the conservation of the rest mass during the cosmological evolution.

We are now ready to find the group-fields in terms of $\theta^{0}$ and $\Lambda$. From the condition $j_{(m)}^{t}=n_{(m)}$ we have that
\begin{equation*}
\theta^{a}=\left[\frac{4R}{\alpha_{(m)}}t+\theta^{0}_{in},0,0,0\right]
\end{equation*}
Considering that equation~(\ref{eqn:dSDJ^a}) reduces to
\[
j_{(m)}^{t}(\partial_{t}\Lambda^{a}_{\;0}+w^{a}_{tb}\Lambda^{b}_{\;0})
=0
\]
we find that $\Lambda^{a}_{\;0}=(1,0,0,0)$ and therefore
\[
\Lambda=\left(\begin{array}{cc}
1 & 0\\ 0 & O
\end{array}\right)
\qquad\textrm{with }O\in \mathrm{SO}(3).
\]
This is not unexpected considering that taking $\alpha_{(s)}=0$ enhance the symmetries of the initial action (\ref{eqn:dSaction}). 

Given the intrinsic homogeneous and isotropic nature of the FRW metric, the spin density was expected to be zero. In general we expect the proposed model to be an optimal framework to study inhomogeneous and$\slash$or anisotropic cosmological spacetimes, where the inhomogeneity and the anisotropy are generated by a primordial spin density.

%%%%%%%%%%%%%%%%%%%%%%%%%%%%%%%%%%%%%%%%%%%%%%%%%%%
\subsubsection{Fluids in Kerr-metric}
\label{sec:Kerr}
%%%%%%%%%%%%%%%%%%%%%%%%%%%%%%%%%%%%%%%%%%%%%%%%%%%
 In this section we consider our fluid description in a Kerr-background:
\[
ds^{2}
=\left(1-\frac{2Mr}{\Sigma}\right)dt^{2}
+2\frac{2Mr}{\Sigma}a\sin^{2}\theta dtd\phi
-\frac{\Sigma}{\Delta}dr^{2}
-\Sigma d\theta^{2}
-\left(r^{2}+a^{2}+\frac{2Mr}{\Sigma}a^{2}\sin^{2}\theta\right)\sin^{2}\theta d\phi^{2}
\]
with
\[
\Delta=r^{2}+a^{2}-2Mr \quad\mbox{and}\quad
\Sigma=r^{2}+a^{2}\cos^{2}\theta.
\]

%The Kerr metric has two Killing vectors, $\partial_{t}$ and $\partial_{\phi}$, therefore\marginpar{what is interpretation?}
%\[
%j^{\nu}\nabla_{[\nu}\left(\frac{F'}{n}j_{\mu]}K^{\mu}\right)
%=j^{\nu}\nabla_{[\nu}\left(\frac{F'}{n}j_{\mu]}\right)K^{\mu}
%+\frac{1}{2}\frac{F'}{n}j^{\nu}j^{\mu}\nabla_{\nu}K_{\mu}
%-\frac{1}{2}nF'\nabla_{\mu}K^{\mu}
%=j^{\nu}\nabla_{[\nu}\left(\frac{F'}{n}j_{\mu]}\right)K^{\mu}
%\]

For the complexity of the equations we will consider the case of a stationary fluid localized on the $\theta=\pi/2$ plane
\[
j_{(i)}^{\mu}=n_{(i)}[u_{(i)}^{t}(r),0,0,u_{(i)}^{\phi}(r)]=n_{(i)}u_{(i)}^{t}(1,0,0,\omega_{(i)})
\]
and with zero vorticity
\[
j_{(m)\mu}=\frac{\alpha_{(m)}n_{(m)}}{4F_{(m)}R}\partial_{\mu}\theta_{0}.
\]
Notice that such a current automatically satisfies the condition $\nabla_{\mu}j_{(i)}^{\mu}=0$, $n_{(i)}$ being function of $r$ only.

Let's solve for dust. From the stationary condition and the expression for the mass current we have
\[
\frac{\alpha_{(m)}}{4R}\partial_{\mu}\theta_{0}=(u_{(m)t},0,0,u_{(m)\phi})
\]
that implies
\[
\theta_{0}(t,\phi)=Et-L\phi
\]
and hence
\[
u_{(m)\mu}=\frac{\alpha_{(m)}}{4R}(E,0,0,-L).
\]
Here the constant $E$ and $L$ are respectively the energy and the angular momentum, per unit of $n$, of the fluid (see appendix~\ref{ap:Keplerian}) once we set $\alpha_{(m)}/(4R)$ to $1$ with an appropriate choice of units.
%Then we have
%\[
%j_{(m)}^{\mu}=n_{(m)}\frac{\alpha_{(m)}E}{4R}\frac{1}{\Delta}\frac{r^{3}+a^{2}r+2Ma(a-l)}{r}(1,0,0,\omega_{(m)})
%\]
%with
The above expression needs to satisfy the normalization condition
\[
u_{(m)}^{\mu}u_{(m)\mu}
=\left(\frac{\alpha_{(m)}E}{4R}\right)^{2}\frac{1}{\Delta}\frac{r^{3}+a^{2}r+2Ma(a-l)}{r}(1+\omega_{(m)}l)
=1,
\]
where we defined $l\equiv L/E$, and where the angular velocity $\omega_{(m)}=u_{(m)}^{\phi}/u_{(m)}^{t}$ is
\[
\omega_{(m)}
=\frac{2Ma+(r-2M)l}{r^{3}+a^{2}r+2Ma(a-l)}
\]
that in return implies that we are restricting to fluids in rings around the rotating center. In appendix~\ref{ap:Keplerian} we will explain how this indeed reproduces asymptotically the expected Keplerian rotational behavior in the galaxies. From now on, until the end of this section each expression is evaluated in $r=r_{f}$, where $r_{f}$ is the radial coordinate of the particular ring under consideration.

As in the previously studied case of dust with zero vorticity we can readily set
\[
\Lambda^{a}_{\;\;0}=e^{a}_{\mu}u_{(m)}^{\mu}
\]
or equivalently,
\begin{eqnarray}
\Lambda_{a0} &=& e_{a}^{\mu}u_{(m)\mu}\nonumber\\
&=&\frac{\alpha_{(m)}E}{4R}
(e_{0}^{t}-e_{0}^{\phi}l,0,0,e_{3}^{t}-e_{3}^{\phi}l)\nonumber\\
&=&\frac{\alpha_{(m)}E}{4R}\left[
\frac{r^{2}+a^{2}-al}{r\sqrt{\Delta}},0,0,\frac{1}{r}(a-l)
\right]_{r=r_{f}}.\label{eqn:dS_K_L_a0}
\end{eqnarray}
Using the equations (\ref{eqn:dSDJ^20}), (\ref{eqn:dSDJ^30})
\begin{equation}\label{eqn:dS_K_b0}
j_{(s)}^{\mu}(\Lambda^{-1})^{b}_{\;\;c}\nabla_{\mu}\Lambda^{c}_{\;\;0}=0 \qquad\textrm{with }b=2,3
\end{equation}
reduces to
\[
(\Lambda^{-1})^{b}_{\;\;1}\left[
\left(j_{(s)}^{t}\frac{M}{r^{2}}-j_{(s)}^{\phi}\frac{a(M+r)}{r^{2}}\right)\Lambda^{0}_{\;\;0}
-j_{(s)}^{\phi}\frac{\sqrt{\Delta}}{r}\Lambda^{3}_{\;\;0}
\right]_{r=r_{f}}
\!\!\!\!\!\!\!\!=0
\]
This equation has two possible solutions: we find $\omega_{(s)}$  or we find $\Lambda_{12}=\Lambda_{13}=0$. Here we study the second case that, as we will see, corresponds to not observe any spin current ($j_{(s)}^{\mu}=0$).

Using the group properties of the Lorentz group
\[
\Lambda_{ai}\Lambda^{a}_{\;\;0}=0 \qquad
\Lambda_{ai}\Lambda^{a}_{\;\;i}=-1 \qquad
\textrm{with }i=2,3 \qquad\textrm{and}\quad
\Lambda_{a2}\Lambda^{a}_{\;\;3}=0
\]
we derive
\[
\Lambda_{02}=\frac{\Lambda_{30}}{\Lambda_{00}}\Lambda_{32} \qquad
\Lambda_{03}=\frac{\Lambda_{30}}{\Lambda_{00}}\Lambda_{33}.
\]\[
\left(\frac{\Lambda_{32}}{\Lambda_{00}}\right)^{2}=1-(\Lambda_{22})^{2} \qquad
\left(\frac{\Lambda_{33}}{\Lambda_{00}}\right)^{2}=1-(\Lambda_{23})^{2} \qquad
\frac{\Lambda_{32}\Lambda_{33}}{(\Lambda_{00})^{2}}=-\Lambda_{22}\Lambda_{23}.
\]
From the above results we can finally write down
\[
\Lambda_{ab}=\left(\begin{array}{cccc}
\Lambda_{00} & 0 & \Lambda_{30}\sin\alpha & \Lambda_{30}\cos\alpha\\
0 & 1 & 0 & 0\\
0 & 0 & \cos\alpha & -\sin\alpha\\
\Lambda_{30} & 0 & \Lambda_{00}\sin\alpha & \Lambda_{00}\cos\alpha
\end{array}\right)
\]
where $\Lambda_{00}$ and $\Lambda_{30}$ are determined by equation~(\ref{eqn:dS_K_L_a0}).

Notice that a rotation by a constant angle generated by $T^{23}$ redefines $\alpha$. This is a symmetry of the action, so $\alpha$ is undetermined and can be set to zero to simplify the above expression
\[
\Lambda_{ab}=\left(\begin{array}{cccc}
\Lambda_{00} & 0 & 0 & \Lambda_{30}\\
0 & 1 & 0 & 0\\
0 & 0 & 1 & 0\\
\Lambda_{30} & 0 & 0 & \Lambda_{00}
\end{array}\right).
\]
Notice that the spin charges are not zero
\[
Q_{23}\propto\alpha_{(s)}\Lambda_{00}
\]
while
\[
j_{(s)}^{\mu}=0 \qquad\Rightarrow\qquad n_{(s)}=0
\]
that satisfies the remaining constraint equations (\ref{eqn:dSDJ^12}), (\ref{eqn:dSDJ^13}). Having $n_{(s)}=0$ simply means that the spin density $n_{(s)}Q_{(s)}$ is zero, that is, the distribution of matter in the Kerr-metric does not show any spin polarization.

\subsubsection{Fluid in a G\"odel Universe}\label{sec:Godel}

A more interesting case is given by the G\"odel universe. Here a fluid with constant mass density $\rho$ and pressure $p$ permeates the universe. The corresponding metric, solution of the Einstein equation with a cosmological constant $\Lambda_{c}$ (the subscript `c' for cosmological is used to differentiate the cosmological constant from the group valued field $\Lambda^{a}_{\phantom{a}b}$) in the presence of the fluid just described, is shown below \cite{Grave:2009zz,Kajari:2004ms}
\begin{equation}
ds^{2}
=dt^{2}-\frac{dr^{2}}{1+\frac{r^{2}}{4a^{2}}}
-r^{2}\left(1-\frac{r^{2}}{4a^{2}}\right)d\phi^{2}+2\frac{r^{2}}{\sqrt{2}a}dtd\phi-dz^{2}.
\end{equation}
Here $a$ is a constant, and the following relations give the complete solution
\[
8\pi(\rho+p)=\frac{1}{a^{2}} \qquad 8\pi p=\Lambda_{c}+\frac{1}{2a^{2}}.
\]

The fluid in a G\"odel universe has a cylindrical symmetry as the one studied in the Kerr metric case. Because of the symmetry we expect a solution similar to the one found in sec.~\ref{sec:Kerr}. The fluid also possesses vorticity oriented along the z-direction. This suggests that, although we can set the spin density to zero ($\alpha_{s}=0$), the spatial group parameters $\theta^{i}$'s become relevant to describe the vorticity of the fluid. Because of the symmetry we expect some redundancy in the solution like for the group field element describing the Kerr metric. Choosing the frame field
\begin{eqnarray}
e^{0}_{\mu}  &=& \left(1,0,\frac{r^{2}}{\sqrt{2}a},0\right)\nonumber\\
e^{1}_{\mu}  &=& \left(0,\frac{1}{\sqrt{1+\left(\frac{r}{2a}\right)^{2}}},0,0\right)\nonumber\\
e^{2}_{\mu}  &=& \left(0,0,r\sqrt{1+\left(\frac{r}{2a}\right)^{2}},0\right)\nonumber\\
e^{3}_{\mu}  &=& \left(0,0,0,1\right)\nonumber
\end{eqnarray}
it is possible to construct a solution where the symmetry is manifest:
\begin{equation}\label{eqn:GLambda}
\Lambda_{ab}=\left(\begin{array}{cccc}
\Lambda_{00} & 0 & 0 & -\Lambda_{30}\\
0 & 1 & 0 & 0\\
0 & 0 & 1 & 0\\
\Lambda_{03} & 0 & 0 & -\Lambda_{33}
\end{array}\right),
\end{equation}
where $\Lambda_{00}=-\Lambda_{33}=\cosh\sigma$, $\Lambda_{03}=-\Lambda_{30}=\sinh\sigma$, and $\sigma$ an unknown function on the spacetime. For any solution $\Lambda_{ab}$, the group valued field $e^{-i\beta T_{12}}\Lambda_{ab}e^{i\beta T_{12}}$, with $\beta$ a constant real parameter, is also a solution. For this reason we can deduce that the only spatial $\theta^{i}$ that contributes to the vorticity is $\theta^{3}$, that is, the one along the z-direction, $T_{30}$ being the only $T_{a0}$ generator that commutes with $T_{12}$. It is then easy to see that, in this condition, the transformation
\[
\Lambda\to e^{-i\beta T_{12}}\Lambda e^{i\beta T_{12}}
\]
indeed leaves invariant the mass-current (\ref{eqn:rdj_m}).

In the present case the current (\ref{eqn:rdj_m}), setting $\theta^{1}=\theta^{2}=0$, reduces to
\begin{eqnarray}
j_{(m)\mu}
&=& \frac{n_{(m)}}{4F_{(m)}}\frac{\alpha_{(m)}}{R}\partial_{\mu}\theta_{0}
-\frac{n_{(m)}}{F_{(m)}}\tr\left[
\frac{\theta^{3}}{2R}\alpha_{(m)}T_{30}\Lambda^{-1}\nabla_{\mu}\Lambda
\right]\nonumber\\
&=& \frac{n_{(m)}}{4F_{(m)}}\frac{\alpha_{(m)}}{R}\left[
\partial_{\mu}\theta_{0}+\theta^{3}(\Lambda^{-1}\nabla_{\mu}\Lambda)_{30}
\right]\nonumber
\end{eqnarray}
Using expression (\ref{eqn:GLambda})
\[
(\Lambda^{-1}\nabla_{\mu}\Lambda)_{30}
=\Lambda_{03}\partial_{\mu}\Lambda_{00}
-\Lambda_{33}\partial_{\mu}\Lambda_{30}
=(\sinh^{2}\sigma-\cosh^{2}\sigma)\partial_{\mu}\sigma
=-\partial_{\mu}\sigma
\]
which yields
\begin{equation}\label{eq:GodelJ}
j_{(m)\mu}
=\frac{n_{(m)}}{4F_{(m)}}\frac{\alpha_{(m)}}{R}\left[
\partial_{\mu}\theta_{0}+\theta_{3}\partial_{\mu}\sigma
\right].
\end{equation}
On the other hand expressions~(\ref{eqn:dSDJ^a}) and (\ref{eqn:dSDJ^30}) give
\[
\partial_{t}\sigma=0 \qquad \partial_{t}\theta_{3}=0.
\]
The remaining equations (\ref{eqn:dSDJ^10}, \ref{eqn:dSDJ^20}, \ref{eqn:dSDJ^12}, \ref{eqn:dSDJ^13}) are identically verified.
From the expression for the co-moving current we have
\begin{eqnarray}
\partial_{t}\theta_{0} &=& \frac{4F_{(m)}R}{\alpha_{m}} \nonumber\\
\partial_{r}\theta_{0}+\theta_{3}\partial_{r}\sigma &=& 0 \nonumber\\
\partial_{\phi}\theta_{0}+\theta_{3}\partial_{\phi}\sigma &=& \frac{r^{2}}{\sqrt{2}a}\frac{4F_{(m)}R}{\alpha_{m}} \nonumber\\
\partial_{z}\theta_{0}+\theta_{3}\partial_{z}\sigma &=& 0 \nonumber
\end{eqnarray}
The solutions to the above system can be found in appendix~\ref{ap:Godel}.

At this point it is easy to show that the fluid has a vorticity
\[
\omega_{\mu\nu}=\nabla_{[\mu}\left(\frac{F_{(m)}}{n_{(m)}}j_{\nu]}\right)
=\frac{\alpha_{(m)}}{4R}\partial_{[\mu}\theta_{3}\partial_{\nu]}\sigma
=\frac{F_{(m)}}{2}\frac{r}{\sqrt{2}a}\delta_{[\mu}^{r}\delta_{\nu]}^{\phi}
\]
and the following energy-momentum tensor
\begin{eqnarray}
T^{\mu\nu} &=&
-\left[F_{(m)}n_{(m)}-F(n_{(m)})\right]g^{\mu\nu}
+\frac{F_{(m)}}{n_{(m)}}j_{(m)}^{\mu}j_{(m)}^{\nu}
\nonumber
\end{eqnarray}
with 
\[
\rho=F=\frac{1}{8\pi}\left(\frac{1}{2a^{2}}-\Lambda_{c}\right) \qquad
p=n_{(m)}F_{(m)}-F=\frac{1}{8\pi}\left(\frac{1}{2a^{2}}-\Lambda_{c}\right).
\]

This example shows how it is possible to describe more general fluids settings involving vorticity. In particular we can notice that (\ref{eq:GodelJ}) is reminiscent of the Clebsch parametrization described in sec.~\ref{sec:ordinary_abelian_fluids}. This is not unexpected because the Clebsch parametrization describes a fluid without spin-density. In this case, indeed it is always possible to choose the $e_{3}^{\mu}$ frame field along the direction of the local axis of rotation of the fluid reducing the number of $\theta^{i}$'s parameters to just $\theta_{3}$ and $\Lambda_{ab}$ to the form (\ref{eqn:GLambda}).

%%%%%%%%%%%%%%%%%%%%%%%%%%%%%%%%%%%%%%%%%%%%%%%%%%%%
\section{Alternative Approaches}
\label{sec:alternative_approaches}
%%%%%%%%%%%%%%%%%%%%%%%%%%%%%%%%%%%%%%%%%%%%%%%%%%%

In this section we explore two alternative approaches to generalize the relativistic spinning fluid action introduced in \cite{Nair:2011mk}. The first is based on a direct use of the Poincar\'e group following \cite{Balachandran:1977ub}. The second approach uses again the de Sitter group and a reduction procedure similarly to section~[\ref{sec:dS_Spinning_fluids}] but in the contest of Mac-Dowell-Mansouri gravity in which the coupling between spin and torsion comes naturally. Also this approach finds its origins in \cite{Balachandran:1977ub} but in a slightly different version described in \cite{Freidel:2006hv} that makes it suited for Mac-Dowell-Mansouri gravity.

One main difference between the models described in this section and the one in section~[\ref{sec:dS_Spinning_fluids}] is that here both models contains only one single current while in the previous formulation we separate the spin and the mass transport into two currents, effectively increasing the degrees of freedom in the dynamics.

\subsection{Fluids Using Poincar\'e Group}
\label{sec:P_Spinning_fluids}

In this section, to find both the mass and spin transport by a fluid we use the full Poincar\'e group. We generalize to fluids in line with the model introduced for relativistic spinning particles in \cite{Balachandran:1977ub}.

Following \cite{Balachandran:1977ub}, the action for spinning fluids introduced in \cite{Nair:2011mk} can be generalized to
\begin{equation}
S=
-\int\det e\, d^{4}x\left[
-F[(u\Lambda^{-1})_{a}e^{a}_{\mu}j^{\mu}]
+i\frac{\lambda}{2}j^{\mu}\tr(T_{12}\Lambda^{-1}\nabla_{\mu}\Lambda)
\right]
\end{equation}
where $u=(1,0,0,0)$, the function $F$ depends on the equation of states and $\lambda$ has the units of angular velocity and represents the ``spin-charge'' of the particles composing the fluid. Notice that, like in the previously described models, $j^{\alpha}$ is conserved, the action being invariant under the right action of a constant element in the $T_{12}$ direction. The variation with respect to $j^{\mu}$ gives
%\[
%\delta_{j}S=
%-\int\det e d^{4}x \delta j^{\mu}\left[
%-F'(u\Lambda^{-1})_{a}e^{a}_{\mu}
%+i\frac{\lambda}{2}\tr(T_{12}\Lambda^{-1}\nabla_{\mu}\Lambda)
%\right],
%\]
%that is,
\[
-F'(u\Lambda^{-1})_{a}e^{a}_{\mu}
+i\frac{\lambda}{2}\tr(T_{12}\Lambda^{-1}\nabla_{\mu}\Lambda)=
0
\]
or equivalently,
\begin{equation}\label{eqn:j}
v_{\mu}=
\frac{i}{F'}\frac{\lambda}{2}\tr(T_{12}\Lambda^{-1}\nabla_{\mu}\Lambda)
\end{equation}
where
\[
v_{\mu}
=(u\Lambda^{-1})_{a}e^{a}_{\mu}
=(\Lambda^{-1})^{0}_{\;a}e^{a}_{\mu}
\]
is the fluid velocity field naturally constructed as boost with respect to the local frame field. The variation of the action with respect to $\Lambda$ yields the spin-precession equation
\begin{equation}\label{eqn:spin-precession}
2F'j^{[a}v^{b]}-\nabla_{\mu}(j^{\mu}S^{ab})=0
\end{equation}
where  $i\lambda\Lambda T_{12}\Lambda^{-1}=S$ is the spin-density.

The Mathisson-Papapetrou equation can be constructed following the same steps as in section~[\ref{sec:dS_Spinning_fluids}]
\begin{equation}\label{eqn:MDM-MP}
j^{\nu}\nabla_{\nu}(F'v_{\mu})-j^{\nu}\nabla_{\mu}(F'v_{\nu})=
\frac{1}{2}\tr(j^{\nu}\nabla_{\nu}S\nabla_{\mu}\Lambda\Lambda^{-1})
+\tr(SR_{\nu\mu})j^{\nu}
\end{equation}
where we multiplied by $j^{\nu}$.

The energy-momentum tensor is given by
\begin{align*}\label{Eq:stresstensor}
T^{\alpha\beta}&=
-\left[
-F[(u\Lambda^{-1})_{b}e^{b}_{\mu}j^{\mu}]
+i\frac{\lambda}{2}j^{\mu}\tr(T_{12}\Lambda^{-1}\nabla_{\mu}\Lambda)
\right]g^{\alpha\beta}
+F'v_{a}e^{a\beta}j^{\alpha}
-\nabla_{\gamma}(j^{\beta}Q^{\gamma\alpha}+j^{\alpha}Q^{\gamma\beta})\nonumber\\
&=-\left(
v_{\mu}j^{\mu}F'
-F
\right)g^{\alpha\beta}
+F'v^{\beta}j^{\alpha}
-\nabla_{\gamma}(j^{\beta}Q^{\gamma\alpha}+j^{\alpha}Q^{\gamma\beta})
\end{align*}
Note that the energy tensor $T^{\alpha\beta}$ is not symmetric because of the appearance of the non-symmetric factor $-F'v^{\beta}j^{\alpha}$.

The main reason for which in general we have to expect a non-symmetric energy-momentum tensor is the presence of a term in the action linear in the frame field: $j^{\mu}v_{a}e^{a}_{\mu}$. Only in the case in which we have an action in terms of the frame field such that only quadratic combinations appear, we will have a symmetric energy-momentum tensor because each quadratic combination of the frame field can be replaced by the full metric. The non-symmetry of the energy-momentum tensor implies\footnote{A second option would be a modification of the gravitational sector that would produce a non-symmetric Einstein tensor. In this paper we shall not address this option.} the presence of a constraint in the theory as is obvious from the Einstein's equation
\[
G^{\alpha\beta}=8\pi T^{\alpha\beta}
\]
Indeed the lhs of the above equation is symmetric by construction which in return implies
\begin{equation}\label{eqn:constraint}
T^{[\alpha\beta]}=-F'v^{[\beta}j^{\alpha]}=0
\end{equation}
implying that $j^{\alpha}$ and $v^{\alpha}$ are proportional!

The first consequence of the constraint (\ref{eqn:constraint}) is that equation (\ref{eqn:spin-precession}) becomes
\begin{equation}\label{eqn:spin-precession+constrain}
\nabla_{\mu}(j^{\mu}S^{ab})=0
\end{equation}
effectively limiting the dynamics. The second consequence comes straightforwardly considering that the parallel-condition
\[
j^{\alpha}=\alpha v^{\alpha},
\]
together with equation (\ref{eqn:j}) and the dispersion relation $j^{\alpha}j_{\alpha}=n^{2}$, imply
\[
n^{2}=
\alpha^{2}v_{\alpha}v^{\alpha}
=\alpha^{2},
\]
that is,
\[
j^{\alpha}=\pm nv^{\alpha}
\]
and
\[
n^{2}=
j^{\alpha}j_{\alpha}=
i\frac{1}{F'}\frac{\lambda}{2}j^{\alpha}\tr(T_{12}\Lambda^{-1}\nabla_{\alpha}\Lambda).
\]
This means that $j^{\alpha}$ is the current of the fluid and that some of the degrees of $\Lambda$ are constrained by the specific choice of $F(n)$. Note that we want to get dynamics with $F(n)$ in the action. To achieve it for mostly minus metric signature, we need to choose the plus sign in the above relation. Notice that the current $j^{\alpha}$ was not previously defined by an equation of motion, its definition in fact comes from the above constraint.

In the presence of our newly found constraint, the Mathisson-Papapetrou equation reduces to
\begin{equation*}
j^{\nu}\nabla_{\nu}(F'v_{\mu})-j^{\nu}\nabla_{\mu}(F'v_{\nu})=
\tr(SR_{\nu\mu})j^{\nu}
\end{equation*}

%%%%%%%%%%%%%%%%%%%%%%%%%%%%%%%%%%%%%%%%%%%%%%
\subsection{Fluid Using de Sitter Group with Torsion}
\label{sec:dS_torsion}
%%%%%%%%%%%%%%%%%%%%%%%%%%%%%%%%%%%%%%%%%%%%%%

Mac-Dowell-Mansouri gravity is a $SO(4,1)$-gauge theory with a term that breaks the symmetry down to the Lorentz group to recover Einstein's Gravity with torsion. In this framework it is natural to introduce particles within the group theoretical approach of \cite{Balachandran:1977ub} as described in \cite{Freidel:2006hv}. We want to generalize this procedure to fluids.

The fluid action will be of the same form of the actions that we described previously:
\begin{equation}\label{eqn:MDM-A}
S
=-\int \mbox{det} \,e\, [ij^{\mu}\tr(Kg^{-1}D_{\mu}g)+F(n)],
\end{equation}
where $g\in SO(4,1)$, $K=c_{0}T_{0}+c_{23}T_{23}$ and
\[
D_{\mu}=\partial_{\mu}+[A_{\mu},\phantom{-}]
\qquad
A_{\mu}
=w_{\mu}^{ab}T_{ab}+\frac{2}{l}e_{\mu}^{a}T_{a}.
\]
The gravitational potential above is the $SO(4,1)$-gauge potential of the Mac-Dowell-Mansouri Gravity where $l$ denotes the de Sitter radius; $l\to\infty$ is the reduction limit of the theory. The variation of the full action, including the gravitational sector, with respect to the frame field $e$ will produce Einstein's equations while the variation with respect to the spin connection $w$ will give us the torsion equation. As we have already seen in the previous section, the presence of the frame field $e$ has to do with the translational part of the Poincar\'e; indeed in the reduction limit the coefficient $c_{0}$ will correspond to the momentum density of the fluid.

The gauge group is broken from the gravity sector to $SO(3,1)$ but this does not reduces the field $g$ that will still have values in $SO(4,1)$. 

Noticing that the right action of $f=\exp{\alpha T_{0}}$ would send an element of the $SO(3,1)$-gauge subgroup into an element of the $SO(4,1)$-gauge group while $f=\exp{\beta T_{23}}$ would leave it in the $SO(3,1)$-gauge subgroup and considering that an $SO(3,1)$-gauge transformation is internal in $SO(3,1)$, we see that it is consistent to study the restriction of the $g$-field to the $SO(3,1)$-subgroup only. The general case of $g\in SO(4,1)$ has a more rich dynamics but here we are going to restrict to the case $SO(3,1)$-valued $g$ that reproduces the expected equations \cite{Freidel:2006hv}.

The conservation of the current follows from the variation of the action (\ref{eqn:MDM-A}) with respect to $g$, giving
\begin{equation}\label{eqn:MDM-DJ=0}
0
=D_{\mu}(j^{\mu}J^{ab})
=\partial_{\mu}(j^{\mu}J^{ab})
+\Gamma_{\mu\gamma}^{\mu}j^{\gamma}J^{ab}
+2j^{\mu}{w_{\mu}}^{a}_{c}s^{cb}
-\frac{1}{2l}j^{\mu}e^{[a}_{\mu}p^{b]}
\end{equation}
where
\[
J=
gKg^{-1}
=c_{0}\Lambda_{0}^{a}T_{a}+c_{23}\Lambda_{2}^{a}\Lambda_{3}^{b}T_{ab}
\equiv p^{a}T_{a}+s^{ab}T_{ab}
\]
The action is also invariant under the right action $g\to gf$, with $f\in SO(3,1)$ and commuting with $K$. It is easy to see that the only commuting elements in $so(4,1)$ with $K$ are the generators $T_{0}$ and $T_{23}$. Such global symmetries imply  the conservation of the current
\begin{equation}\label{Dj=0}
\nabla_{\mu}j^{\mu}=0.
\end{equation}
Using (\ref{Dj=0}), the previous expression (\ref{eqn:MDM-DJ=0}) reduces to the spin precession equation
\begin{equation}\label{Eq:SPdS}
j^{\mu}\nabla_{\mu}J^{ab}
-\frac{1}{2l}j^{\mu}e^{[a}_{\mu}p^{b]}
=0
\end{equation}
The variation with respect to the current $j^{\alpha}$ produces
\[
j_{\mu}=-i\frac{n}{F'}\tr[Kg^{-1}D_{\mu}g]
\]
This, following the usual procedure already defined in section~[\ref{sec:dS_Spinning_fluids}] give the Mathisson-Papapetrou equation for fluids
\begin{align*}\label{Eq:MPdS}
j^{\nu}\nabla_{\nu}\left(\frac{F'}{n}j_{\mu}\right)
-j^{\nu}\nabla_{\mu}\left(\frac{F'}{n}j_{\nu}\right)
&=-i\tr\left(j^{\nu}\nabla_{\nu}J\nabla_{\mu}gg^{-1}\right)
-2ij^{\nu}\tr\left(JR_{\nu\mu}\right)\nonumber\\
&=-i\frac{1}{2l}\tr\left(j^{\mu}e^{[a}_{\mu}p^{b]}T_{ab}\nabla_{\mu}gg^{-1}\right)
-2ij^{\nu}\tr\left(JR_{\nu\mu}\right)
\end{align*}

%%%%%%%%%%%%%%%%%%%%%%%%%%%%%%%%%%%%%%%%%%%
\section{Conclusion}
\label{sec:conclusion}
%%%%%%%%%%%%%%%%%%%%%%%%%%%%%%%%%%%%%%%%%%%

In this work, we presented a formal model to compute the analogue of the Mathisson-Papapetrou equation for fluids which are relevant in strong gravitational backgrounds\footnote{The fact that it is in the realm of strong gravitational effect can be understood by the fact that the spin coefficient is small unless it is coupled to gravity strongly. Similar reasoning holds for Torsion.}. Our approach relies on the group theoretical study started with works of various authors, as mentioned in the introduction. In sections~[\ref{sec:dS_Spinning_fluids}] and~[\ref{sec:alternative_approaches}], we have presented several prospective models based on different prescriptions, reproducing the full dynamics of a fluid with spin in terms of both the mass and spin transport. 

We also present two more models in section~3, which are different and new solutions to the same problem. These two alternative models are based on the same line of thinking, but in two different contexts: the use of the Poincar\'e group instead of the de~Sitter group and its consequences, and how to include torsion, that could be relevant given that the spin-density of the fluid is one of the degrees of freedom considered in the model.

With the discovery of QGP in the laboratory, we have recently seen a booming interest on such strongly coupled field theoretic systems. Various properties of such system are only possible to be described by AdS/ CFT like dualities for which one needs to do weak coupling calculation in the gravity side which most of the time introduces a black hole horizon and so on. But the scope of application of such dualities are still limited and we hope that our present alternative studies of fluid dynamics might help us to side-track this general limitation.

As per future directions, there are several avenues that we can take to proceed. Though on the negative side of this formalism we note that it's quite difficult and ad hoc to incorporate the viscosity and other dissipative effects, but the fact that the formulation allows us to study various general backgrounds in the presence of fluids with spin-densities makes us hopeful for further astrophysical studies. One possible imminent direction would be to study anistotrpic and{\slash}or inhomogeneous cosmological models where there is the presence of a primordial vorticity. Such studies may have important consequences on our understanding of the CMB. Other possible directions include a generalization of the present model to include viscosity.
\vspace{1cm}

%%%%%%%%%%%%%%%%%%%%%%%%%%%%%%%%%%%%%
\begin{center}
\textbf{Acknowledgements}
\end{center}
%%%%%%%%%%%%%%%%%%%%%%%%%%%%%%%%%%%%%

We are indebted to V.P. Nair and the referees of our article for some very useful discussions throughout this project and during publication. The work of D.C. is supported by Templeton Foundation grant ID 21531 and D.S. is supported in part by US National Science Foundation grant PHY-0855582 and Lehman College CUNY Science Fellowship.

\vspace{5cm}
%%%%%%%%%%%%%%%%
\appendix
%%%%%%%%%%%%%%%%
\begin{center}
\large{\textbf{Appendix}}
\end{center}
\section{de Sitter Group Generators}
\label{appendixC}
%%%%%%%%%%%%%%%%%%%%%%%%%%%%%%%%%%%%%%%%%%%%%%%%%%%%%

In this appendix we list the convention used in this article for the representation of the Lia algebra associated with the de Sitter group.

The generators of $so(4,1)$ are constructed, similarly to $so(3,1)$, starting the four-dimensional Dirac's gamma matrices
\[
\{\gamma^{a},\gamma^{b}\}=2\eta^{ab}
\]
with the addition of $\gamma^{5}$. Such algebra can be rewritten in a shortened notation as
\[
\{\gamma^{A},\gamma^{B}\}=2\eta^{AB}
\]
where $A$ takes values $\{0,1,2,3,5\}$ and $\eta^{AB}=\{1,-1,-1,-1,-1\}$.

Similarly to the case of the Lorentz group the de Sitter group is generated from Lie algebra generator defined as follows
\begin{eqnarray*}
T_{AB}=\frac{1}{8}\left[\gamma_A,\gamma_B\right]\qquad\mbox{with}\qquad T_{A5}=-T_{5A}\equiv T_A
\end{eqnarray*}

The commutation relations between the Lie Generators are
\[
2[T_{AB},T_{CD}]
=\eta_{BC}T_{AD}-\eta_{AC}T_{BD}-\eta_{BD}T_{AC}+\eta_{AD}T_{BC},
\]
that is,
\[
2[T_{ab},T_{cd}]
=\eta_{bc}T_{ad}-\eta_{ac}T_{bd}-\eta_{bd}T_{ac}+\eta_{ad}T_{bc}
\]\[
2[T_{ab},T_{c}]
=\eta_{bc}T_{a}-\eta_{ac}T_{b} \qquad
2[T_{a},T_{c}]
=T_{ac}
\]

The normalized traces of the Lie generators are given by
\[
-2\tr[T_{AB}T_{CD}]
=\frac{1}{2}(\eta_{AC}\eta_{BD}-\eta_{AD}\eta_{BC}).
\]

%%%%%%%%%%%%%%%%%%%%%%%%%%%%%%%%%%%%%%%%%%%%%%%%%%%%%
\section{Keplerian Rotational behavior}
\label{ap:Keplerian}
%%%%%%%%%%%%%%%%%%%%%%%%%%%%%%%%%%%%%%%%%%%%%%%%%%%%%

To simplify the problem we assumed from the beginning circular orbits. In this appendix we will also consider a non-zero radial component in the current and only at the end we will set such a component to zero to recover circular orbits.

For the dust case under discussion we set
\[
j_{(m)\mu}
=nu_{\mu}
=n(E,u_{r},0,-L).
\]
The constraints on $u_{\mu}$ come from the zero vorticity condition and its normalization\footnote{We have already shown that the current is trivially conserved for $u_{r}=0$.}:
\[
\nabla_{[\mu}u_{\nu]}=0 \qquad u_{\mu}u^{\mu}=1.
\]
Multiplying the zero vorticity condition by $2u^{\mu}$ and using the normalization condition we find
\[
0
=2u^{\mu}\nabla_{[\mu}u_{\nu]}
=u^{\mu}\nabla_{\mu}u_{\nu}.
\]
Let's start from $u^{\mu}\nabla_{\mu}u_{r}$. All the other components will be automatically zero once we set $u_{r}=0$. From the normalization condition
\[
1
=\frac{r^{3}+a^{2}r+2Ma^{2}}{r\Delta}E^{2}
-\frac{4Ma}{r\Delta}EL
-\frac{r-2M}{r\Delta}L^{2}
-\frac{\Delta}{r^{2}}u_{r}^{2}
\]
we have
%\[
%2u_{r}\partial_{r}u_{r}
%=\left(2\frac{r}{\Delta}-2r^{2}\frac{r-M}{\Delta^{2}}\right)\left[
%\frac{\Delta}{r^{2}}u_{r}^{2}
%\right]
%+\frac{r^{2}}{\Delta}\left[
%\left(
%-\frac{1}{r}-2\frac{r-M}{\Delta}
%\right)\left(
%1+\frac{\Delta}{r^{2}}u_{r}^{2}
%\right)
%+\frac{(3r^{2}+a^{2})E^{2}-L^{2}}{r\Delta}
%\right]
%\]

\[
u^{r}\partial_{r}u_{r}
=-\frac{1}{2}\left(
\frac{r}{\Delta}
-4r^{2}\frac{r-M}{\Delta^{2}}
\right)\left[
\frac{\Delta^{2}}{r^{4}}u_{r}^{2}
\right]
-\frac{1}{2}\left[
\frac{-3r^{2}-a^{2}+4Mr+(3r^{2}+a^{2})E^{2}-L^{2}}{r\Delta}
\right]
\]
Therefore we can easily expand
%\[
%u^{\mu}\nabla_{\mu}u_{r}
%=-\frac{1}{2}\left(
%\frac{r}{\Delta}
%-4r^{2}\frac{r-M}{\Delta^{2}}
%\right)\left[
%\frac{\Delta^{2}}{r^{4}}u_{r}^{2}
%\right]
%-\frac{1}{2}\left[
%\frac{-3r^{3}-a^{2}r+4Mr^{2}+(3r^{3}+a^{2}r)E^{2}-L^{2}r}{r^{2}\Delta}
%\right]+
%\]\[
%+\frac{L^{2}}{r^{2}\Delta^{2}}\left[
%r^{3}-4Mr^{2}+4M^{2}r-a^{2}M
%\right]
%-\frac{E^{2}M}{r^{2}\Delta^{2}}\left[
%r^{4}+2a^{2}r^{2}+a^{4}-4Ma^{2}r
%\right]
%+\frac{EL}{r^{2}\Delta^{2}}2aM(3r^{3}-4Mr^{2}+a^{2}r)
%\]
%
%\[
%u^{\mu}\nabla_{\mu}u_{r}
%=-\frac{1}{2}\left(
%\frac{r}{\Delta}
%-4r^{2}\frac{r-M}{\Delta^{2}}
%\right)\left[
%\frac{\Delta^{2}}{r^{4}}u_{r}^{2}
%\right]
%-\frac{1}{2r^{2}\Delta^{2}}[
%-3r^{5}-4a^{2}r^{3}+10Mr^{4}
%-a^{4}r+6Ma^{2}r^{2}
%-8M^{2}r^{3}
%\]\[
%+(3r^{5}+4a^{2}r^{3}-6Mr^{4}
%+a^{4}r-2Ma^{2}r^{2}
%)E^{2}
%-(r^{3}+a^{2}r-2Mr^{2})L^{2}+
%\]\[
%+L^{2}\left[
%2r^{3}-8Mr^{2}+8M^{2}r-2a^{2}M
%\right]
%-E^{2}M\left[
%2r^{4}+4a^{2}r^{2}+2a^{4}-8Ma^{2}r
%\right]
%+4aM(3r^{3}-4Mr^{2}+a^{2}r)EL
%]
%\]

\[
u^{\mu}\nabla_{\mu}u_{r}
=-\frac{1}{2}\left(
\frac{r}{\Delta}
-4r^{2}\frac{r-M}{\Delta^{2}}
\right)\left[
\frac{\Delta^{2}}{r^{4}}u_{r}^{2}
\right]
-\frac{1}{2r^{2}\Delta^{2}}[
-3r^{5}-4a^{2}r^{3}+10Mr^{4}
-a^{4}r+6Ma^{2}r^{2}
-8M^{2}r^{3}
\]\[
+(3r^{5}+4a^{2}r^{3}-8Mr^{4}
+a^{4}r-6Ma^{2}r^{2}
-2a^{4}M+8M^{2}a^{2}r
)E^{2}
-(-r^{3}+a^{2}r+6Mr^{2}-8M^{2}r+2a^{2}M)L^{2}+
\]\[
+4aM(3r^{3}-4Mr^{2}+a^{2}r)EL
]
\]
In this appendix we are interested in showing that the theoretically expected Keplerian rotational behaviour is reproduced, therefore we will consider only the asymptotic behaviour, for which we can take $a\to0$:
%\[
%u^{\mu}\nabla_{\mu}u_{r}
%=-\frac{1}{2}\left(
%\frac{r}{\Delta}
%-4r^{2}\frac{r-M}{\Delta^{2}}
%\right)\left[
%\frac{\Delta^{2}}{r^{4}}u_{r}^{2}
%\right]
%-\frac{1}{2r^{2}\Delta^{2}}[
%-3r^{5}+10Mr^{4}
%-8M^{2}r^{3}
%\]\[
%+(3r^{5}-8Mr^{4})E^{2}
%-(-r^{3}+6Mr^{2}-8M^{2}r)L^{2}
%]
%\]
%whereas we can use the normalization condition
%\[
%r^{3}-2Mr^{2}
%=r^{3}E^{2}
%-(r-2M)L^{2}
%\]
%to eliminate $E$. We finally get
\[
u^{\mu}\nabla_{\mu}u_{r}
=-\frac{1}{2}\left(
\frac{r}{\Delta}
-4r^{2}\frac{r-M}{\Delta^{2}}
\right)\left[
\frac{\Delta^{2}}{r^{4}}u_{r}^{2}
\right]
-\frac{1}{2r^{2}\Delta^{2}}[
-4Mr^{3}(r-2M)
+4r(r-3M)(r-2M)L^{2}
]
\]
Setting $u_{r}=0$ to simplify our condition (in this way we are going to assume only circular orbits), we have
\[
L=\sqrt{\frac{Mr_{f}^{2}}{r_{f}-3M}}
\]
and hence
\[
E=\sqrt{\frac{(r_{f}-2M)^{2}}{r_{f}(r_{f}-3M)}}
\]
that indeed reproduces the expected Keplerian behaviour of the rotation curve for galaxies
\[
\omega
=\frac{g^{\phi\phi}(-L)}{g^{tt}E}
=\sqrt{\frac{M}{r_{f}^{3}}}.
\]

\section{Solving for the Clebsch-parameters in the G\"odel Universe}\label{ap:Godel}
In this section we shall find a solution for the system
\begin{eqnarray}
\partial_{t}\theta_{0} &=& \frac{4F_{(m)}R}{\alpha_{m}} \nonumber\\
\partial_{r}\theta_{0}+\theta_{3}\partial_{r}\sigma &=& 0 \nonumber\\
\partial_{\phi}\theta_{0}+\theta_{3}\partial_{\phi}\sigma &=& \frac{r^{2}}{\sqrt{2}a}\frac{4F_{(m)}R}{\alpha_{m}} \nonumber\\
\partial_{z}\theta_{0}+\theta_{3}\partial_{z}\sigma &=& 0 \nonumber
\end{eqnarray}
From taking the $z$-derivative of the first two equations
\begin{eqnarray}
\partial_{r}\sigma\partial_{z}\theta_{3}-\partial_{r}\theta_{3}\partial_{z}\sigma &=& 0 \nonumber\\
\partial_{\phi}\sigma\partial_{z}\theta_{3}-\partial_{\phi}\theta_{3}\partial_{z}\sigma &=& 0 \nonumber
\end{eqnarray}
and noticing that the determinant of the system
\[
-\partial_{r}\sigma\partial_{\phi}\theta_{3}+\partial_{r}\theta_{3}\partial_{\phi}\sigma
=\frac{r}{\sqrt{2}a}\frac{2F_{(m)}R}{\alpha_{m}}
\]
is proportional to the rotation vector associated to the fluid, we directly obtain the expected result that the group parameter are $z$-independent
\[
\partial_{z}\theta_{3}=0 \qquad \partial_{z}\sigma=0 \qquad \partial_{z}\theta_{0}=0.
\]

We can use the following ansatz
\[
\theta_{0}=\frac{4F_{(m)}R}{\alpha_{m}}\left(t+\frac{r^{2}}{\sqrt{2}a}\right) \qquad \partial_{\phi}\sigma=0
\]
which follows looking at the $\phi$ component of $u_{\mu}$. From here it is easy to deduce that
\[
\theta_{3}=-a\phi \qquad \sigma=\frac{r^{2}}{\sqrt{2}a^{2}}.
\]
Notice that the coefficient above were chosen to make the parameters of the right dimension.

Like in the case of the Clebsch parametrization we can redefine $\theta_{0}$ and $\theta_{3}$
\[
\theta_{0}\to\theta_{0}+\gamma(\sigma) \qquad
\theta_{3}\to\theta_{3}-\frac{d\gamma}{d\sigma}
\]
where $\gamma$ is a generic function of $\sigma$.


\begin{thebibliography}{99}

\bibitem{Bistrovic:2002jx} 
  B.~Bistrovic, R.~Jackiw, H.~Li, V.~P.~Nair and S.~Y.~Pi,
  ``NonAbelian fluid dynamics in Lagrangian formulation,''
  Phys.\ Rev.\ D {\bf 67}, 025013 (2003)
  [hep-th/0210143].
  %%CITATION = HEP-TH/0210143;%%

\bibitem{Jackiw:2004nm} 
  R.~Jackiw, V.~P.~Nair, S.~Y.~Pi and A.~P.~Polychronakos,
  ``Perfect fluid theory and its extensions,''
  J.\ Phys.\ A A {\bf 37}, R327 (2004)
  [hep-ph/0407101].
  %%CITATION = HEP-PH/0407101;%%

\bibitem{Nair:2011mk} 
  V.~P.~Nair, R.~Ray and S.~Roy,
  ``Fluids, Anomalies and the Chiral Magnetic Effect: A Group-Theoretic Formulation,''
  arXiv:1112.4022 [hep-th].
  %%CITATION = ARXIV:1112.4022;%%

\bibitem{Balachandran:1977ub} 
  A.~P.~Balachandran, S.~Borchardt and A.~Stern,
  ``Lagrangian and Hamiltonian Descriptions of Yang-Mills Particles,''
  Phys.\ Rev.\ D {\bf 17}, 3247 (1978).
  %%CITATION = PHRVA,D17,3247;%%
  A.~P. Balachandran, G. Marmo, B.-S. Skagerstam, A. Stern, 
  ``Gauge Symmetries and Fibre Bundles: Applications to Particle Dynamics'',
  Lecture Notes in Physics, Berlin Springer Verlag, vol. 188, 1983.

\bibitem{Wong:1970fu} 
  S.~K.~Wong,
  ``Field and particle equations for the classical Yang-Mills field and particles with isotopic spin,''
  Nuovo Cim.\ A {\bf 65}, 689 (1970).
  %%CITATION = NUCIA,A65,689;%%

\bibitem{Mathisson:1937zz} 
  M.~Mathisson,
  ``Neue mechanik materieller systemes,''
  Acta Phys.\ Polon.\  {\bf 6}, 163 (1937).
  %%CITATION = APPOA,6,163;%%

\bibitem{Corinaldesi:1951pb} 
  E.~Corinaldesi and A.~Papapetrou,
  ``Spinning test particles in general relativity. 2.,''
  Proc.\ Roy.\ Soc.\ Lond.\ A {\bf 209}, 259 (1951).
  %%CITATION = PRSLA,A209,259;%%

\bibitem{Grave:2009zz} 
  F.~Grave, M.~Buser, T.~Muller, G.~Wunner and W.~P.~Schleich,
  %``The Godel universe: Exact geometrical optics and analytical investigations on motion,''
  Phys.\ Rev.\ D {\bf 80}, 103002 (2009).
  %%CITATION = PHRVA,D80,103002;%%
  %3 citations counted in INSPIRE as of 14 Feb 2014

\bibitem{Kajari:2004ms}
  E.~Kajari, R.~Walser, W.~P.~Schleich and A.~Delgado,
  ``Sagnac effect of Godel's universe,''
  gr-qc/0404032.
  %%CITATION = GR-QC/0404032;%%
  %2 citations counted in INSPIRE as of 14 Feb 2014
  
\bibitem{Freidel:2006hv} 
  L.~Freidel, J.~Kowalski-Glikman and A.~Starodubtsev,
  ``Particles as Wilson lines of gravitational field,''
  Phys.\ Rev.\ D {\bf 74}, 084002 (2006)
  [gr-qc/0607014].
  %%CITATION = GR-QC/0607014;%%

\bibitem{Andersson:2006nr}
  N.~Andersson and G.~L.~Comer,
  ``Relativistic fluid dynamics: Physics for many different scales,''
  Living Rev.\ Rel.\  {\bf 10}, 1 (2005)
  [gr-qc/0605010].
  %%CITATION = GR-QC/0605010;%%
  
\bibitem{Rubin:1970zza} 
  V.~C.~Rubin and W.~K.~Ford, Jr.,
  ``Rotation of the Andromeda Nebula from a Spectroscopic Survey of Emission Regions,''
  Astrophys.\ J.\  {\bf 159}, 379 (1970).
  %%CITATION = ASJOA,159,379;%%  
  %reference article: \begin{verbatim}http://adsabs.harvard.edu/doi/10.1086/150317\end{verbatim}

  
\end{thebibliography}
\end{document}